\documentclass[11pt,a4paper]{article}
\usepackage[margin=1in]{geometry}

\usepackage[natbibapa]{apacite}
\usepackage{hyperref}
\usepackage{amsmath}
\usepackage{graphicx}
\usepackage{esvect}
\usepackage{amssymb}
\numberwithin{equation}{section}
\usepackage{float}
\usepackage{todonotes}
\usepackage{tabularx}
\usepackage{longtable}
\usepackage{pdflscape}
\usepackage{booktabs}
\usepackage{setspace}
\usepackage{multirow}
\usepackage[ruled,vlined]{algorithm2e}

\SetKwComment{Comment}{$\triangleright$\ }{} 

\usepackage{float}
\usepackage{longtable}  

\usepackage{comment}

\usepackage{caption} 
\usepackage{makecell} 

\usepackage{array} 
\newcolumntype{P}[1]{>{\centering\arraybackslash}p{#1}}
\newcolumntype{M}[1]{>{\centering\arraybackslash}m{#1}}

\usepackage{glossaries}
\makeglossaries 

\newacronym[plural={AMRs},\glsshortpluralkey={AMRs}]{amr}{AMR}{autonomous mobile robot}

\newacronym{abswp}{ABSWP}{autonomous block stacking warehouse problem}

\newacronym{upmp}{UPMP}{unit-load pre-marshalling problem}

\newacronym{mupmp}{MUPMP}{multibay unit-load pre-marshalling problem}

\newacronym{rmp}{RMP}{remarshalling problem}

\newacronym{mip}{MIP}{mixed-integer programming}

\newacronym{ram}{RAM}{random-access memory}

\newacronym[plural={AGVs},\glsshortpluralkey={AGVs}]{agv}{AGV}{automated guided vehicle}

\newacronym{slam}{SLAM}{simultaneous localization and mapping}

\newacronym{lidar}{LIDAR}{light imaging, detection and ranging}

\newacronym{rfid}{RFID}{radio-frequency identification}

\newacronym{gtin}{GTIN}{global trade item number}

\newacronym[plural={SKUs},\glsshortpluralkey={SKUs}]{sku}{SKU}{stock keeping unit}

\newacronym{wyswyg}{WYSWYG}{what-you-see-is-what-you-get}

\newacronym[plural={CPS},\glsshortpluralkey={CPS}]{cps}{CPS}{cyber-physical system}

\newacronym{bbd}{BBD}{best before dates}

\newacronym{fifo}{FIFO}{first-in first-out}

\newacronym{lifo}{LIFO}{last-in first-out}

\newacronym[plural={RMFS},\glsshortpluralkey={RMFS}]{rmfs}{RMFS}{robotic mobile fulfillment system}

\newacronym{bim}{BIM}{building information model}

\newacronym{tsp}{TSP}{traveling salesperson problem}

\newacronym{it}{IT}{information technology}

\newacronym[plural={I/O-points},\glsshortpluralkey={I/O-points}]{I/O-point}{I/O-point}{input/output-point}

\newacronym[plural={I-points},\glsshortpluralkey={I-points}]{I-point}{I-point}{input-point}

\newacronym[plural={o-points},\glsshortpluralkey={O-points}]{O-point}{O-point}{output-point}

\newacronym{scc}{SC}{single command}

\newacronym{dcc}{DC}{dual command}

\newacronym{mcc}{MC}{multi command}

\newacronym{mdp}{MDP}{markov decision process}

\newacronym{pomdp}{POMDP}{partially observable markov decision process}

\newacronym{slap}{SLAP}{storage location assignment problem}

\newacronym{cpu}{CPU}{Central Processing Unit}

\newacronym{api}{API}{application programming interface}

\newacronym{ni}{NI}{no information}

\newacronym{pi}{PI}{product information}

\newacronym{ii}{II}{item information}

\newacronym{cb}{CB}{class-based}

\newacronym{rb}{RB}{random-based}

\newacronym{col}{COL}{closest open location}

\newacronym{nn}{NN}{nearest-neighbor}

\newacronym{sl}{SL}{shortest leg}

\newacronym{coi}{COI}{cube order per index}

\newacronym{tob}{TOB}{turnover-based}

\newacronym[plural={KPIs},\glsshortpluralkey={KPIs}]{KPI}{KPI}{key performance indicator}

\newacronym{mcts}{MCTS}{monte carlo tree search}

\newacronym{dos}{DOS}{duration-of-stay}

\newacronym{crp}{CRP}{container relocation problem}

\newacronym{asrs}{AS/RS}{automated storage and retrieval system}

\newacronym{vrp}{VRP}{vehicle routing problem}

\newacronym{fcfs}{FCFS}{first-come first-serve}

\newacronym{lwr}{LWR}{longest waiting retrieval}

\newacronym{nr}{NR}{nearest retrieval}

\newacronym{nsr}{NSR}{nearest storage/retrieval}

\newacronym{bfifo}{BFIFO}{batch first-in, first-out}

\newacronym{fefo}{FEFO}{first-expiry, first-out}

\newacronym{aco}{ACO}{ant colony optimization}

\newacronym{cpmp}{CPMP}{container pre-marshalling problem}

\newacronym[plural={CSP},\glsshortpluralkey={CSP}]{csp}{CSP}{coil shuffling problem}

\newacronym{bg}{BG}{bad-good}

\newacronym{bb}{BB}{bad-bad}

\newacronym{gg}{GG}{good-good}

\newacronym{gb}{GB}{good-bad}

\newacronym{gx}{GX}{good-good and good-bad}

\newacronym{bx}{BX}{bad-good and bad-bad}

\newacronym[plural={FMCG},\glsshortpluralkey={FMCG}]{fmcg}{FMCG}{fast-moving consumer good}

\newacronym{cp}{CP}{constraint programming}

\newacronym{wms}{WMS}{Warehouse Management System}
\newacronym{erp}{ERP}{Enterprise Resource Planning}

\usepackage[utf8]{inputenc}
\usepackage[english]{babel}

\graphicspath{ {./images/} }

\title{Sorting multibay block stacking storage systems\thanks{This work was partly funded by the Deutsche Forschungsgemeinschaft (DFG, German Research Foundation) through the Research Training Group 2193 (Project number: 276879186).}} 

\author{Jakob Pfrommer, Thomas Bömer, Daniyar Akizhanov, Anne Meyer}
\author{ Jakob Pfrommer\footnote{Corresponding author, TU Dortmund University, Leonhard-Euler-Stra{\ss}e 5, 44227 Dortmund, Germany, jakob.pfrommer@tu-dortmund.de, ORCID: \url{https://orcid.org/0000-0003-2492-0621}}, Thomas Bömer \footnote{TU Dortmund University, Leonhard-Euler-Stra{\ss}e 5, 44227 Dortmund, Germany, thomas.boemer@tu-dortmund.de, ORCID: \url{https://orcid.org/0000-0003-4979-7455}}, Daniyar Akizhanov \footnote{Rice University, 6100 Main St, Houston, TX 77005, USA, da44@rice.edu}, Anne Meyer\footnote{TU Dortmund University, Leonhard-Euler-Stra{\ss}e 5, 44227 Dortmund, Germany, anne2.meyer@tu-dortmund.de, ORCID: \url{https://orcid.org/0000-0001-6380-1348}}}
\date{}

\begin{document}
\maketitle

\begin{abstract}
Autonomous mobile robots (AMRs) are increasingly used to automate operations in intralogistics. One crucial feature of AMRs is their availability, allowing them to operate 24/7. 
This work addresses the multibay unit load pre-marshalling problem, which extends pre-marshalling from a single bay to larger warehouse configurations with multiple bays. Pre-marshalling leverages off-peak time intervals to sort a block stacking warehouse in anticipation of future orders. 
These larger warehouse configurations require not only the minimization of the number of moves but also the consideration of distance or time when making sorting decisions. Our proposed solution for the multibay unit load pre-marshalling problem is based on our two-step approach that first determines the access direction for each stack and then finds a sequence of moves to sort the warehouse. In addition to adapting the existing approach that integrates a network flow model and an extended A* algorithm, we additionally present an exact constraint programming approach for the second stage of the problem-solving process. 
The results demonstrate that the presented solution approach effectively enhances the access time of unit loads and reduces the sorting effort for block stacking warehouses with multiple bays. 
\end{abstract}

\section{Introduction}
\label{sec:intro}
\begin{figure}[!ht]
    \centering
    \includegraphics[width=\textwidth]{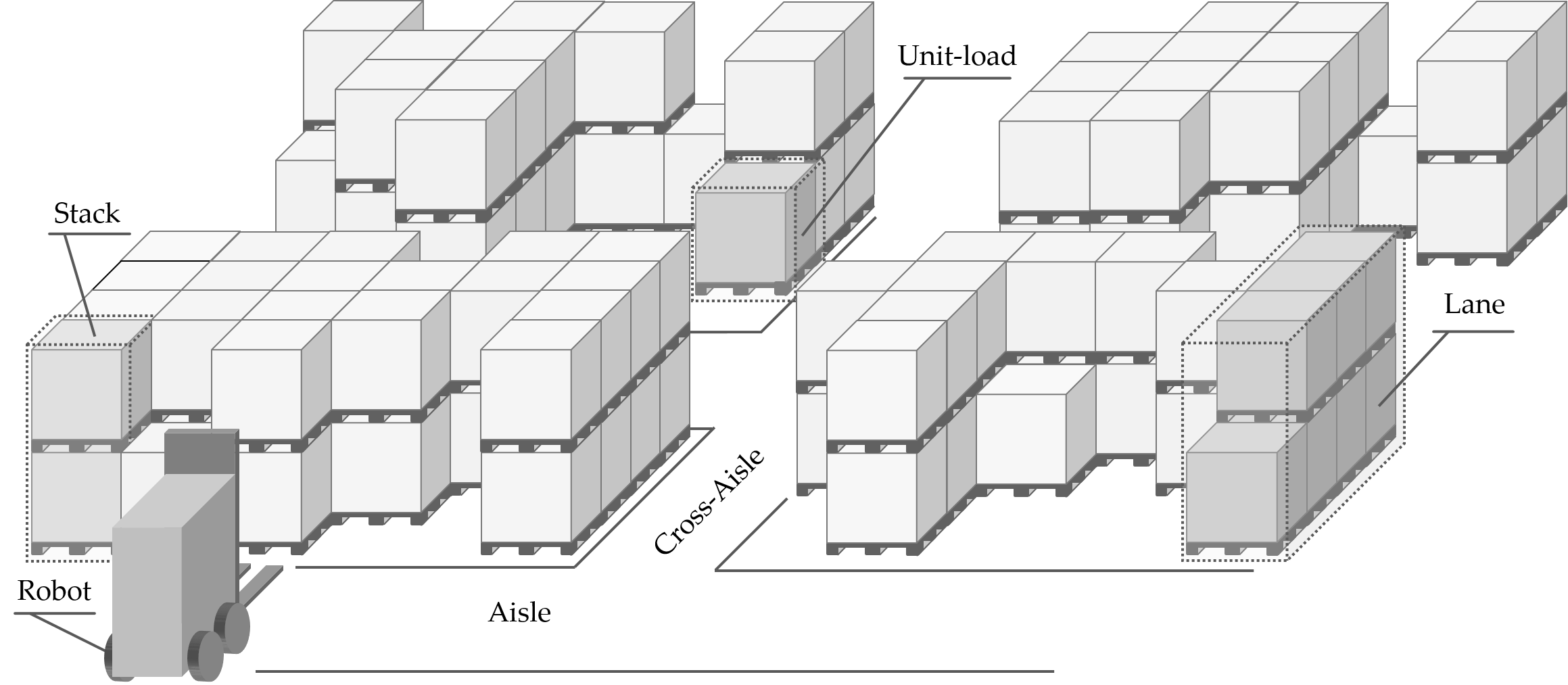}
    \caption{Example of a block stacking warehouse with a robotic forklift} 
    \label{fig:block_storage_example}
\end{figure}

Block stacking storage systems are a very simple and widely used type of warehouse. Unit loads like pallets, boxes, containers, or movable shelves are placed on the ground and may be stacked on top of another. The advantage of block storage is that it does not require high investments in infrastructure, is highly flexible, and easily scalable. Figure \ref{fig:block_storage_example} shows an example of a block stacking warehouse with four bays separated by aisles for traveling. Each bay contains unit loads that are placed in lanes. Dependent on the available access directions and the properties of the unit load, access to the outermost stacks from up to four cardinal directions is possible.

The use of \glspl{amr} for material handling allows to increase the storage density by  
utilizing a shared storage policy and rearrangement operations that are difficult to implement by human operators. The reason is the high availability of \glspl{amr}, also during off-peak hours and accurate information on the storage system state. For example, \glspl{amr} could rearrange the storage during idle time and prepare it for 
a future demand sequence that may be derived from a production schedule or truck loading time slots. In \cite{pfrommer2022solving}, we introduced this problem as \gls{upmp}. This work is the continuation of this paper extending the problem to a larger multibay setup and hence to a much wider range of applications. Extending the scope to multiple bays  
requires not only a highly efficient solution approach but also additional features like the consideration of travel times/distances of the \glspl{amr}. 

The main contributions of this paper are:
\begin{enumerate}
    \item We formally introduce the \gls{mupmp} with up to four access directions.
    \item We show that our two-stage solution approach for the \gls{upmp} can be applied to solve the \gls{mupmp} with little modifications.
    \item We further introduce an optimal \gls{cp} approach for the second stage to evaluate our heuristic tree search approach.
    \item We show that the approaches can solve real-world size instances and publish the benchmark datasets.
\end{enumerate}

This paper is structured in the following main sections.  
We commence this paper with an introduction of the \gls{mupmp} in Section \ref{sec:mupmp} followed by a brief literature review in Section \ref{sec:related_work}. Subsequently, we introduce the extended solution approach in Section \ref{sec:solution} and conduct a series of experiments presented in Section \ref{sec:experiments}. Finally, we provide a summary and outlook in Section \ref{sec:conclusion}.


\section{The Multibay unit load Pre-marshalling Problem}\label{sec:mupmp}
The \gls{mupmp} is about sorting a block stacking warehouse that may contain multiple bays until the resolution of all blockage based on the future retrieval time of each unit load. Similar to the \gls{upmp} for a single bay, we assume that the \glspl{amr} are idle and that no unit loads are entering or leaving the storage system. The warehouse consists of $N$ given bays that are interconnected via aisle space for traveling. Each bay $n \in\{1,...,N\}$ can be accessed from one or up to four access directions and is filled with a set of unit loads with a known position and retrieval priority group. Because a fully shared storage policy is applied, \glspl{sku} can be placed anywhere and mixed in lanes as well as stacks. The retrieval priority group is derived based on a known (e.g., planned truck arrivals or production schedule) or predicted (e.g., based on material supply precedence graph or demand forecast) future retrieval time. We define unit loads as \textit{blocking} when they have a higher retrieval group compared to the unit loads that are placed underneath or behind. These unit loads with a lower retrieval group will be retrieved before the unit load with a higher retrieval group. In block storage, placing a unit load at higher tiers requires a unit load to be located underneath. Only the top unit load can be accessed directly. When filling up or accessing a bay, the available access directions must be considered to prevent empty gaps. In terms of \gls{amr} specifications, we assume that they cannot reach over a stack in front and can only carry a single unit load. 
\par
Whereas in case of the \gls{upmp}, the goal was solely to find a minimum number of unit-load moves to sort a bay such that there is no blockage, in case of the \gls{mupmp}, the move time or distance must also be considered due to the much greater time/distance differences between the moves. Minimizing the total move time allows pre-marshalling to be conducted in short time intervals, thereby ensuring long operational availability. Hence, our goal for solving the \gls{mupmp} is to minimize the travel time/distance for the pre-marshalling process that consists of the total travel as well as the handling times (i.g. pick-up and fine-positioning). 
To do so, we aim for the minimal total loaded move distance for the minimal number of moves. Only distances travelled while an \gls{amr} is loaded are taken into account. This allows an allocation of the moves to multiple \glspl{amr}. Because multibay block stacking warehouses are usually operated by multiple vehicles, it is necessary to split and allocate the resulting sequence of moves to the \glspl{amr}. 
The allocation of moves is the scope of future work.

\begin{figure}[!ht]
    \centering
    \includegraphics[width=\textwidth]{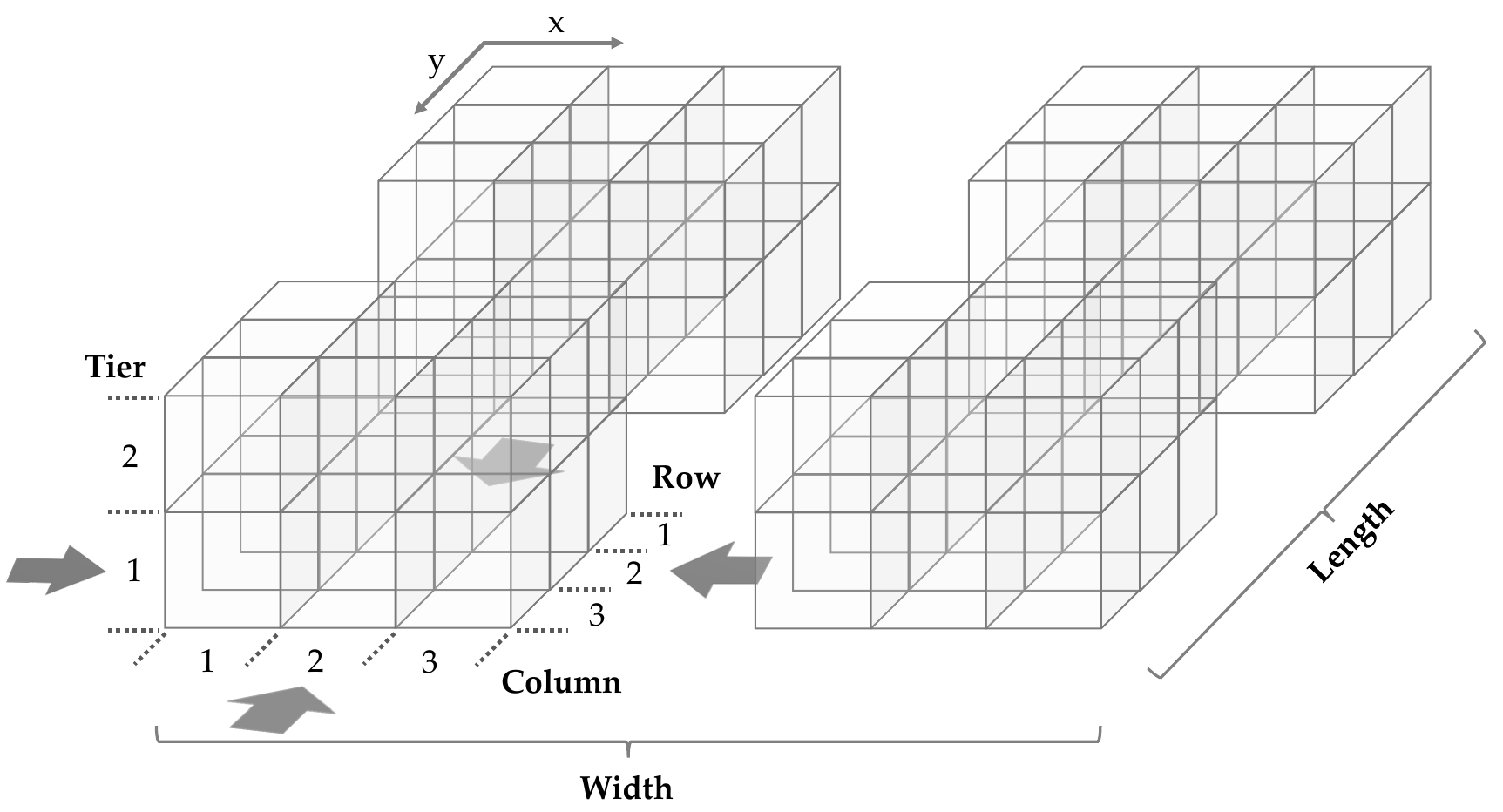}
    \caption{State representation with four bays of the multibay block stacking warehouse} 
    \label{fig:multi_block_storage_example}
\end{figure}

Figure \ref{fig:multi_block_storage_example} shows the state representation of a block stacking warehouse with four bays arranged as a square. Within a bay, the storage locations are represented as a three-dimensional grid with columns $i \in\{1,...,I\}$, rows $j \in\{1,...,J\}$, and tiers $t \in\{1,...,T\}$ that is filled with unit loads of priority groups $g_{ijt} \in\{1,...,G\}$. Assuming rectangular warehouse footprints with length $L$ and width $W$, we also use a two-dimensional global $x$-$y$ coordination system to determine the position of bays, storage locations, and access points in the warehouse. The access points $p \in\{1,...,P\}$ are located in the aisle space in front of the bays and specify the stacks that can be accessed directly. This representation with access points is more detailed than just defining the allowed access directions, as also partial access via one access direction can be modeled, which may be necessary due to pillars, for example. An \gls{mupmp}-instance is fully defined by a list of bays, where each bay contains $(I,J,T,G,g_{ijt} \in G^{I\cdot J\cdot T},P)$. 

\begin{figure}[!ht]
    \centering
    \includegraphics[width=\textwidth]{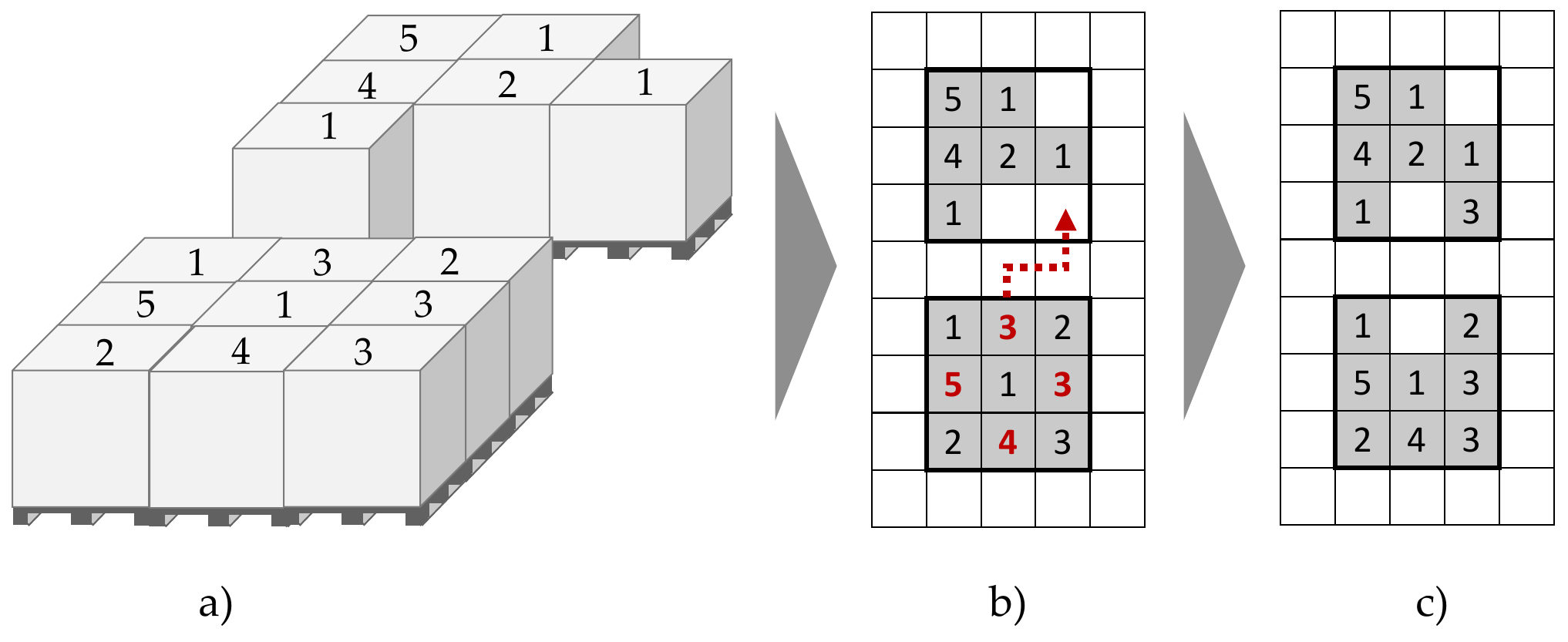}
    \caption{Example configuration in (a), top view of the grid-based layout and red arrow for a possible move in (b) and solution of the \gls{mupmp} in (c)} 
    \label{fig:multi_shuffling_example}
\end{figure}

The example in \ref{fig:multi_shuffling_example} (a) shows a warehouse configuration containing two square bays of the same size with $I=3$, $J=3$, and $T=1$ that can both be accessed from all four directions. It is filled with unit loads of retrieval groups $g_{ijt}$ from one to five that are specified on top of each load. 
\ref{fig:multi_shuffling_example} (b) switches the perspective to the top view with white tiles for empty/aisle space and dark gray tiles for unit loads. The bold red digits are the unit loads that are blocking the center unit load of the bottom bay. A red-dotted arrow indicates a possible move to resolve all blockage. Relocating one of the other blocking unit loads would also be possible, but it requires a longer transport distance. \ref{fig:multi_shuffling_example} (c) shows the resulting sorted configuration of the \gls{mupmp}. 

The \gls{mupmp} increases the scope to a variety of storage applications considering not only a single bay but larger multibay warehouses. Hence, the \gls{mupmp} is a generalization of the \gls{upmp} in \cite{pfrommer2022solving}. In the case of bays that can only be accessed from one access direction, it can be reduced to the \gls{cpmp} and is thus also NP-hard \citep{casertaContainerRehandlingMaritime2011}. Bays with more than once access direction require further investigation on the complexity.



\section{Related work}
\label{sec:related_work}
Our literature review presents related publications focusing on rearrangement activities for multibay setups as well as the minimization of travel time. For an overview of single-bay pre-marshalling literature, we refer to \cite{pfrommer2022solving}.

\subsection{Multiple bays}
In maritime transportation, the rearrangement of containers within multiple bays is known as the \gls{rmp}. Similar to pre-marshalling, the goal is to sort the scattered containers based on their retrieval priority with a minimum number of moves until all blockings are resolved \cite{casertaContainerRehandlingMaritime2011}. Different problem definitions exist for \gls{rmp}, which the following literature review shows. 
\par
An early publication by \citeauthor{kim1998re} \cite{kim1998re} deals with the \gls{rmp} and converting the configuration of a current bay into a target configuration. Instead of the exact position of each container, the bay configuration is defined simply as the number of containers per group that are located within each bay. The group is a collection of containers going to the same destination on the same vessel. The problem is divided into two sub-problems: (1) bay matching and move planning and (2) sequencing the moving tasks. In the first step, dynamic programming and an approach referred to as "transportation algorithm" are used to determine the flow and movement between bays to reach the target configuration. Second, the task sequencing problem is modeled as a \gls{tsp} with precedence constraints to consider space availability at the destination bay, with the goal of minimizing travel time. Because the exact positions of the containers are neglected in the bay configuration, the rehandling of containers within a bay is not addressed in this work.
\par
Later publications take exact container positions into account and investigate the intra-block \gls{rmp} where containers are sorted based on categories and retrieval priorities. In successive publications, \citeauthor{kang2006planning} \cite{kang2006planning} and \citeauthor{choe2011generating} \cite{choe2011generating} investigate the \gls{rmp} where, starting from a source, bay containers are relocated to an empty target bay so that no rehandling is necessary either during the rearrangement process or later in the ship loading process. Interference between multiple cranes is taken into account. Both publications present a similar two-phase solution approach that couples a simulated annealing algorithm for finding a target configuration and partial order graph with crane scheduling to determine a minimum-time re-marshalling plan. 
\par
\citeauthor{park2009planning} \cite{park2009planning} study the \gls{rmp} in a container terminal with two cranes. Whereas one crane is used for loading containers from the storage yard on the vessel (seaside), the second crane is used to carry in containers from external trucks (landside). The goal of their solution is to bring containers from the landside of the storage yard to the seaside where the ship loading happens and, at the same time, sort the containers based on their retrieval priority. Therefore, the authors propose a two-stage algorithm: First, potential stacks where the containers will be moved are selected heuristically. 
Second, an algorithm is used to conduct an alternating search to solve the two sub-problems of determining target slots that are assigned to each container as well as the order of container movements. 
\par
Further approaches deal with rearrangement activities to improve the location of containers whenever one of the cranes is temporarily idle (called "housekeeping" \cite[p 148]{kemme2012design} or "iterative rescheduling"), where the re-marshalling jobs that most reduce future loading times are selected with the objective of minimizing current delay times as well as the overall makespan \cite{choe2015crane}. 
Besides re-marshalling before a ship is loaded (export terminal), re-marshalling activities may also be performed after unloading a ship (import terminal) before containers are retrieved (e.g., to external trucks). The works of
\citeauthor{van2013evaluating}\cite{van2013evaluating} and \citeauthor{covicRemarshallingAutomatedContainer2017} \cite{covicRemarshallingAutomatedContainer2017} investigate this type of re-marshalling problem where container departure times are reported by a truck announcement system.
\par
The presented research for multiple bays deals with a variety of use-case-specific problems (e.g., moving containers to an empty target bay) using relaxing assumptions to simplify the problems. We could not find an extension of the \gls{cpmp} to the \gls{rmp}, possibly because cranes are typically installed permanently for a single block (here: bay) and cannot move between them.

\subsection{Minimization of crane times}
Besides publications that specifically address rearrangement operations between multiple bays, there are also solution approaches of the \gls{cpmp} considering crane times. Instead of minimizing moves, \citeauthor{parreno2020minimizing} \cite{parreno2020minimizing} address the minimization of crane times for the pre-marshalling of containers. The authors propose an integer linear model that builds upon \cite{parreno-torresIntegerProgrammingModels2019} as well as a branch and bound algorithm, which is set up as an iterative deepening search. Besides the lower bound from \cite{tanakaBranchBoundApproach2019}, a novel lower bound incorporating the travel time, as well as a heuristic algorithm for an upper bound for the number of moves, are also introduced. The heuristic algorithm is executed at each iteration of the search, and the algorithm stops when the depth level reaches the upper bound. The novel time-based lower bound extends the lower bound for the number of moves. Each move is multiplied by a minimum time that consists of the travel time unloaded and loaded, crane lowering and lifting times, and twistlock time. Further variants of the time-based lower bound also consider the position of the blocking container and the highest positions they can be moved to. The time-based lower bound is used as a secondary tie-breaking criterion. Overall, the branch and bound algorithm outperforms the integer programming models in terms of solved instances and runtime. In a recent publication, \citeauthor{parreno2022beam} \cite{parreno2022beam} propose a beam search algorithm that is able to find all previously found optimal and more improved solutions (not necessarily optimal). For the tree search procedure, the authors use several tie-breaking criteria based on the move type (\gls{bb}, \gls{bg}, \gls{gb} and \gls{gg}; see \cite{pfrommer2022solving}), develop a global evaluation heuristic to obtain the most promising nodes, and present new dominance rules for eliminating nodes that cannot result in a better solution.
\par
The research on the minimization of crane times by \citeauthor{parreno2022beam} \cite{parreno2022beam} uses a lower bound based on minimum travel times multiplied by the number of moves. In single-bay scenarios, this is a feasible approach. However, in larger multibay warehouses, using a similar lower bound based on the minimum move time would not be an effective approach.

\section{Solving the MUPMP}
\label{sec:solution}

In this section, we give an overview and briefly describe our adapted two-step solution approach for the \gls{mupmp} based on a network flow model and a tree search procedure that has been presented in \cite{pfrommer2022solving}. 

Our main extensions are:
\begin{itemize}
    \item A pre-processing step to determine the access points for each bay and calculate the distance matrix.
    \item Multiple outputs when solving the network flow model for each bay applying the advanced lower bound as an additional tie-breaking rule. 
    \item An improved tree search procedure
    \begin{itemize}
        \item with an incremental calculation of the lower bound heuristics, 
        \item new tie-breaking rule that includes the total loaded move distance, 
    \end{itemize}
    \item An exact \gls{cp} approach to validate the results of the tree search procedure.
\end{itemize}

The process flow of our two-step approach is shown in Figure \ref{fig:diagram_approach_updated}. 

\subsection{Two-step Approach}

\begin{figure}[ht]
    \centering
    \includegraphics[width=9cm]{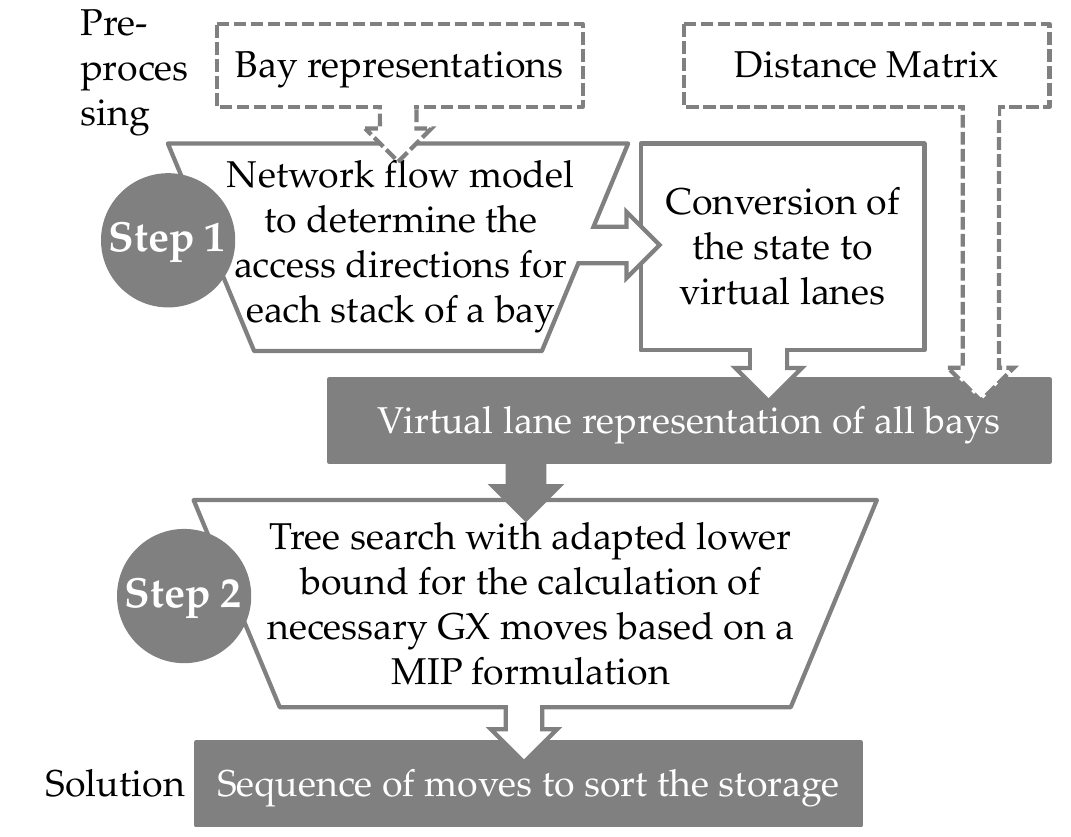}
    \caption[Illustration of the main components of our extended solution approach.]{
    Illustration of the main components of our extended solution approach based on \cite{pfrommer2022solving}.} 
    \label{fig:diagram_approach_updated}
\end{figure}

We start this section by giving an overview and briefly describing the steps of our solution approach based on a network flow model and a tree search procedure that have been presented in \cite{pfrommer2022solving}. The process flow of our two-step approach is shown in Figure \ref{fig:diagram_approach_updated}. 
Beforehand, a pre-processing step converts the warehouse representation into a list of bays with access points that indicate from which aisle locations stacks are accessible. Based on a graph that is placed in the grid-based layout, we calculate a static distance matrix with the distances between all access points. 

In the first step, we use the network flow model to determine the access directions for each stack. The objective is to minimize the blocking unit loads. Based on the access directions, we can transform the configuration into a state representation of virtual lanes. Each virtual lane consists of a sequence of 1 to $\max (I,J)$ stacks in a row/column that can be accessed one after the other. The access direction for each virtual lane is defined by an access point, and the length of the lane may be different. In the second step, we use the virtual lane representation to find a solution for sorting the warehouse with a minimum number of moves. Therefore, we present an adapted tree search procedure as well as a \gls{cp} approach. 

In the following subsections, we explain our solution approach in detail and focus on the newly added adaptations as well as the additional pre- and post-processing steps.

\subsection{Input}
\begin{figure}[ht]
    \centering
    \includegraphics[width=9.5cm]{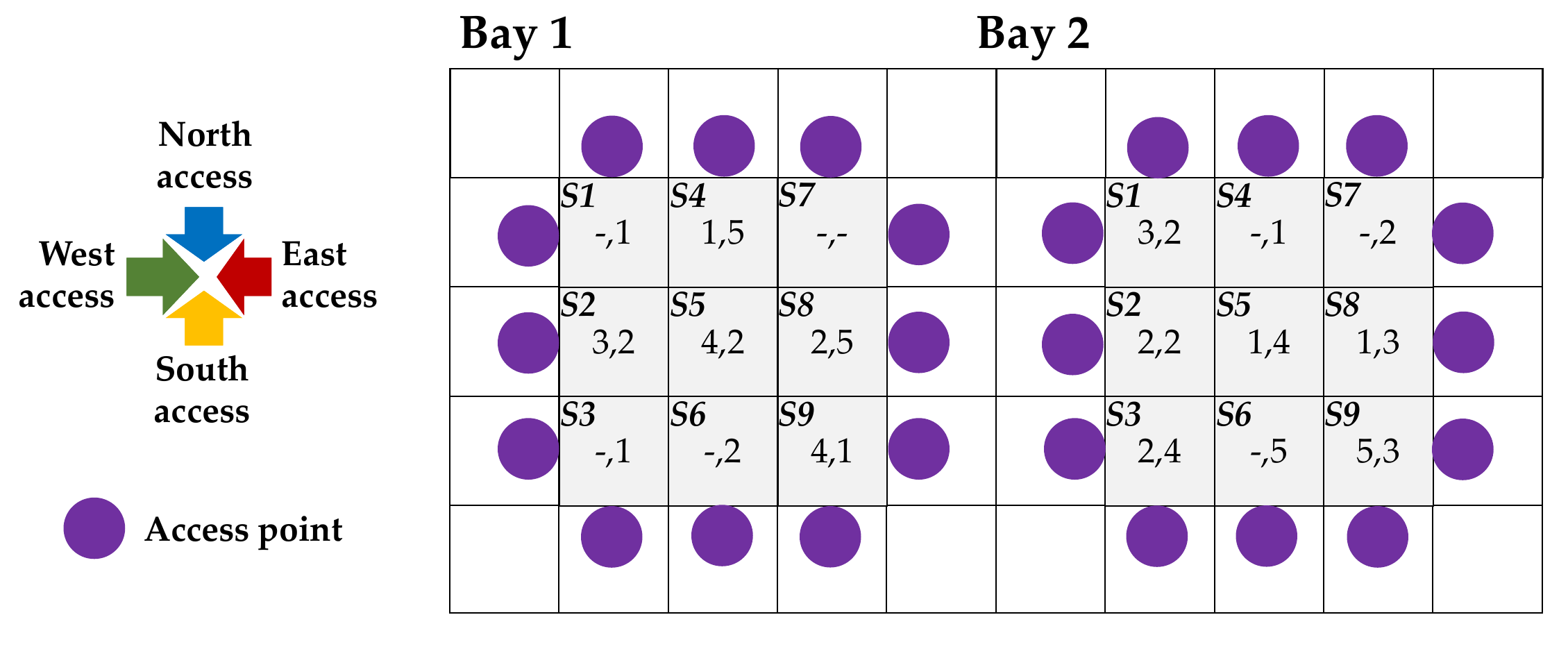}
    \caption{Example of a warehouse configuration with two bays and possible access points.}
    \label{fig:example_state_rep_ap}
\end{figure}

In the beginning, we add the required inputs. A grid-based warehouse is represented as a list of bays including the access points and directions to the respective bay as shown in Figure \ref{fig:example_state_rep_ap}. The purple dots indicate possible access points from aisle space to the bay. This example allows access from all four cardinal directions. 
Next, we lay a graph into the aisle space tiles and calculate all distances between the access points of all bays via breadth-first-search. 
A downside of using such a static distance matrix between all access points is that possible shortcuts through storage space in case of low fill levels are not considered.

\subsection{Step 1: Access direction fixing}
\label{subsec: Step 1: Access direction fixing}
In this step, we want to define the access direction for each stack. This is necessary to make the application of our lower bound valid and also because we want to avoid crossing \glspl{amr} when accessing a bay. Fixing the access directions at the beginning of our approach may lead to additional moves, as discussed in \cite{pfrommer2022solving}. To calculate the access direction for each stack that minimizes the number of misplaced unit loads, we use the network flow model from \cite{pfrommer2022solving}. The result is used to determine a virtual lane representation that defines how each stack is accessed via a defined access point. 
In our previous version, we obtained a single optimal solution in case many symmetrical solutions were available. Now, we want to utilize situations in which many symmetrical solutions are available by returning multiple optimal solutions for each bay. We then compare up to 10 solutions with our more advanced lower bound heuristic that also considers the priority groups of blocking unit loads and open locations (see \cite{pfrommer2022solving}). 

\begin{figure}[ht]
    \centering
    \includegraphics[width=\textwidth]{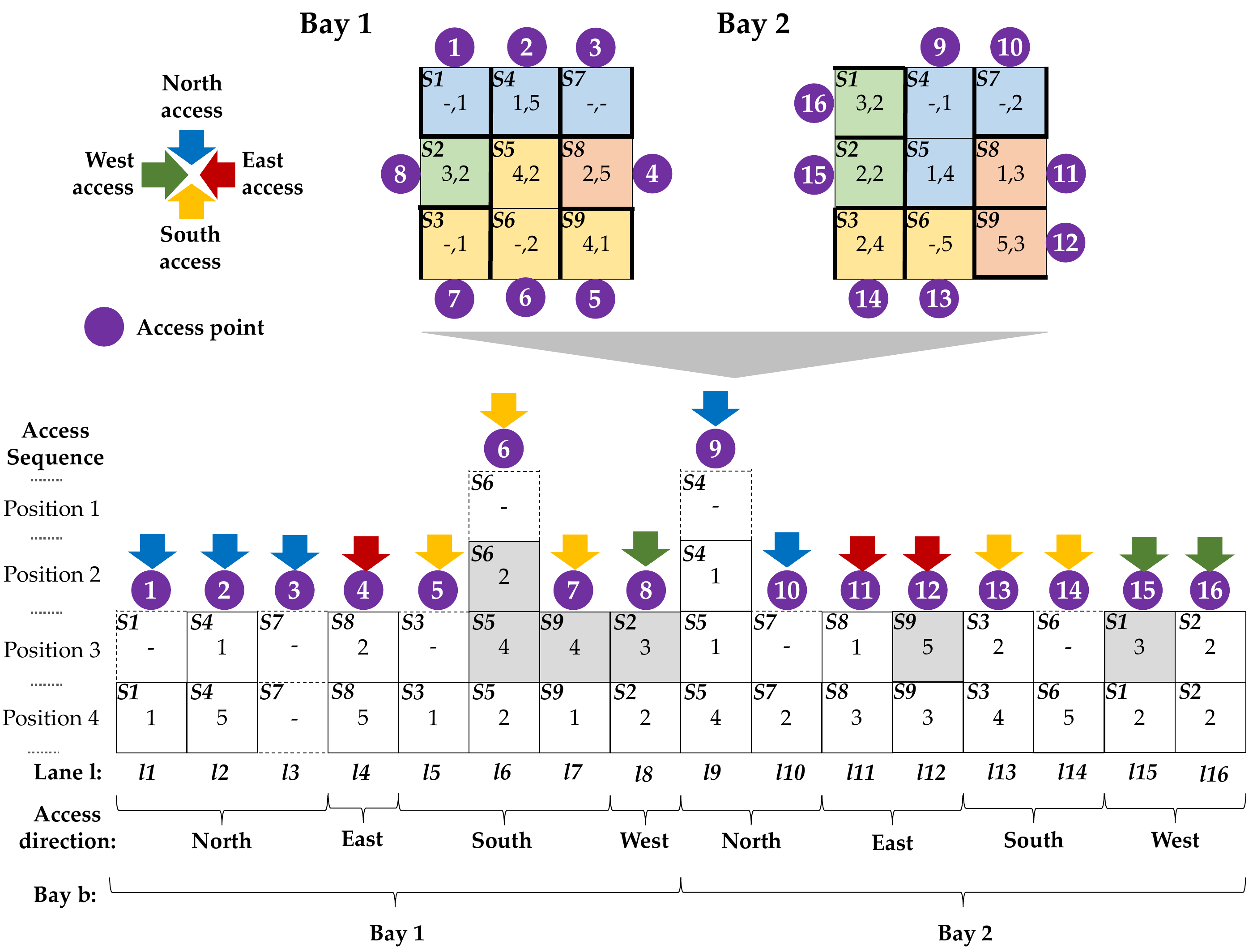}
    \caption[Resulting virtual lanes for the example.]{Resulting virtual lanes for the example from \ref{fig:example_state_rep_ap}. 
    Thick black lines in bay 1 and bay 2 illustrate the virtual lanes. These lanes are transformed into a representation of one access direction. }
    \label{fig:example_virtual_lanes}
\end{figure}

Figure \ref{fig:example_virtual_lanes} gives an example with two bays that produce a list of virtual lanes. The two bays on top of the figure show an exemplary result of the network flow model for the example from Figure \ref{fig:example_state_rep_ap}. Thick black lines and colors illustrate the virtual lanes with the respective access directions and access points. At the bottom of the figure, we show the final virtual lane representation that is used throughout the following procedure to determine the pre-marshalling moves.

\subsection{Step 2 - Alternative 1: Tree search}
\label{subsec: Step 2 - Alternative 1: Tree search}

The new implementation of the A* algorithm differs in some details from our previous work. We introduce a new tie-breaking rule that considers the total loaded move distance for popping nodes out of the priority queue and a faster incremental calculation of the lower bound heuristics (which is not visible in the algorithm). 
\par
Algorithm \ref{astar_mod} shows the modified algorithm. 
After initialization, the main search loop runs until a solution with no blocking unit loads is found, until the open list is empty and no solution exists, or until the time limit has been reached. 
In each iteration of the loop, a node $n$ that minimizes the cost estimation $f(n) = g(n) + h(n)$ is popped off the open list, where $g(n)$ is the number of moves already performed at node $n$ and $h(n)$ the lower bound estimating the number of moves until a blockage-free configuration is reached. If there are many nodes with the same cost estimation $f(n)$, we use a new tie-breaking rule for node selection that is explained in the subsequent paragraphs. 
The only difference in calculating successor nodes is that we incrementally calculate and update the total loaded move distance. 
A successor node is pushed to the open list and added to an f- and distance-dictionary, if there is not already a node with the same configuration in the f-dictionary that has a smaller cost estimation or the same cost estimation but a smaller total loaded move distance. Our procedure guarantees an exact solution for the minimization of moves. Since the travel distance is considered only for tie-breaking, the resulting total loaded move distance is not necessarily optimal for the minimal number of moves. As long as the processing time for handling exceeds the travel time for each move, we expect good results for the total 
processing time as well.

\begin{algorithm}[ht]
\SetAlgoLined
 \SetKwFunction{Fastar}{astar}
 \SetKwProg{Fn}{Function}{:}{}
 \Fn{\Fastar{root}}{
 open $\leftarrow$ \textsc{priority\_queue}()\Comment*[r]{new tie-breaking rules}
 f\_lookup $\leftarrow$  \textsc{dict}()\;
 distance\_lookup $\leftarrow$ \textsc{dict}()\Comment*[r]{new distance lookup}
 closed $\leftarrow$  $\emptyset$\;
 \textsc{push}(open, root)\;
 f\_lookup[root] $\leftarrow$ $f($root$)$\;
 distance\_lookup[root] $\leftarrow$ $0$\;
 \While{open $>$ $0$ and time $<$ timeout}{
  $n$ $\leftarrow$ \textsc{pop}(open)\;
  closed $\leftarrow$ closed\: $\bigcup$\: \{$n$\}\;
  \If{blocking($n$) $= 0$}{
   \KwRet $n$\;
  }
 successors $\leftarrow$ \textsc{branching}($n$)\Comment*[r]{new extended branching over all bays}
  \For{$s$ $\in$ successors}{
    \If{$s$ $\notin$ closed and ($s$ $\notin$ f\_lookup or f\_lookup[$s$] $>$ f($s$) or (f\_lookup[$s$] $=$ f($s$) and distance\_lookup[$s$] $>$ total\_loaded\_distance($s$)))} {
     \textsc{push}(open, $s$)\;
     f\_lookup[$s$] $\leftarrow$ f($s$)\;
     distance\_lookup[$s$] $\leftarrow$ total\_loaded\_distance($s$)\;
  }
  }
 }
 }
\caption{Modified A* algorithm from \cite{pfrommer2022solving}.}
\label{astar_mod}
\end{algorithm}

\textbf{Tie-breaking rules with travel distance:} 
When selecting the next move, our algorithm compares the different rules in a lexicographic order. The applied rules for tie-breaking are the f-value (total cost estimation), h-value (lower bound heuristics), and the total loaded move distance $dist$ (aggregated from the root to this node). The distance between two virtual lanes is defined as the length of the shortest path between their respective access points. 
The travel time within each virtual lane is neglected (it can be added by counting the empty tiles). 
We build different combinations based on these rules. In our approach, we always select the possible moves with the lowest f-value first. Because in many cases (especially large warehouses) there are still plenty of options with the same minimum f-value, we use the h-value as the first, and the total loaded move distance $dist$ as the second tie-breaker: $f\_h\_dist$.

\textbf{Incremental lower bound calculation}:
The A* search is further optimized by incremental lower bound calculation, which is not visible in Algorithm \ref{astar_mod}. If the configuration and lower bound of a given node are known, calculating the lower bound of its successor can be simplified. The estimated number of required moves based on the \gls{bx} and the \gls{gx} moves can be computed faster if information about its predecessor is known. The number of misplaced unit loads can be changed only in the lanes impacted by a move. Therefore, the difference in misplaced unit loads within these lanes is therefore also the total difference in misplaced unit loads in the entire warehouse. The \gls{bx} moves based on the minimum number of misplaced unit loads per virtual lane, can also be updated simply by comparing it to the number of misplaced unit loads in the virtual lanes affected by the most recent move. The calculation of additional \gls{gx} moves uses a supply-and-demand model. The change in supply and demand is dependent only on the change in the virtual lanes affected by the most recent move. 
To calculate the difference, the supply and demand of the two virtual lanes affected by the move must to be determined for the parent node and the successor. The difference in supply and demand is calculated by subtracting the parent's supply and demand from the successor's supply and demand. Based on the supply and demand differences, we calculate the cumulative incremental supply and demand, deriving the demand surplus difference. 
We then add the demand surplus difference to the demand surplus of the parent node to obtain the new demand surplus of the successor. 
The following steps to derive the \gls{gx} moves are the same.

\subsection{Step 2 - Alternative 2: Constraint programming}
\label{subsec: Step 2 - Alternative 2: Constraint programming}
To demonstrate the performance of the A* algorithm, we compare its solution to a \gls{cp} approach. 
We employ a \gls{cp} approach proposed by \cite{jimenez2023constraint}. \citeauthor{jimenez2023constraint} solve the \gls{cpmp} with the minimum number of moves. The core concept involves constructing a \gls{cp} model that allows a limited number of moves, starting from a lower bound of moves. If no solution emerges after a complete search, an additional move is incorporated into the model until a solution is identified. 
The authors introduce multiple model versions. We opt for CP5 as it proved to be the most efficient. In the subsequent sections, we detail only the relevant parts of the model that underwent modifications to solve the \gls{mupmp}. The complete model is described in appendix \ref{appendix_constraint programming}:
\par
Each stack of containers in the original model equates to a virtual lane, and each tier equates to a position in the access sequence of the virtual lane (see section \ref{subsec: Step 1: Access direction fixing}). To ensure comparability, the variable names used are matched to the original model. Let $\bar s$ denote the number of virtual lanes and $\bar t$ represent the number of positions in each virtual lane $s$. Accordingly, $\mathcal{S} = \{1, . . ., \bar s\}$ is the set of virtual lanes and $\mathcal{T}_{s} = \{1, . . .,\bar t\}$ is the set of positions for virtual lane $s \in \mathcal{S}$.  
A slot ($s$,$t$) is defined by a virtual lane $s \in \mathcal{S}$ and position $t  \in \mathcal{T}_{s}$. 
Let $\bar k$ be the permissible number of moves in the model. $\mathcal{K}$ is the set of stages, defined as $\mathcal{K} = \{1, . . ., \bar k\}$. Stage 0 signifies the initial layout. $\mathcal{K}^{0} = \{0,1, . . ., \bar k\}$ is the set of stages, including the initial layout.
Further, the variables $z^{k}_{s,t}$ and $y^{k}_{i,j}$ are introduced.

\begin{align*}
z^{k}_{s,t} &=
    \begin{cases}
      1 & \text{If a unit load is removed from slot ($s$,$t$) during stage $k$}\\
      0 & \text{Otherwise}
    \end{cases}
\\
&\hspace*{3.8em} \forall s \in \mathcal{S}, \forall t \in \mathcal{T}_{s} , \forall k \in \mathcal{K}
\\
y^{k}_{s,t} &=
    \begin{cases}
      1 & \text{If a unit load is moved to slot ($s$,$t$) during stage $k$ } \\
      0 & \text{Otherwise}
    \end{cases}
\\
&\hspace*{3.8em} \forall s \in \mathcal{S}, \forall t \in \mathcal{T}_{s}, \forall k \in \mathcal{K}
\end{align*}

Let $d_{s, s'}$ be the distance between virtual lane $s \in \mathcal{S}$ and $s' \in \mathcal{S}$.
The total loaded move distance is minimized for the minimum number of moves. The objective function is defined as:
\begin{equation}
min
    \displaystyle\sum_{k \in \mathcal{K}}
    \displaystyle\sum_{s \in \mathcal{S}}
    \displaystyle\sum_{s' \in \mathcal{S}}
    \displaystyle\sum_{t \in \mathcal{T}_{s}}
    \displaystyle\sum_{t' \in \mathcal{T}_{s'}}
    z^{k}_{s,t} \cdot y^{k}_{s',t'} \cdot d_{s,s'}
\end{equation}
Let $c^{UB}$ be the objective value of the solution of the A* algorithm. To restrict the search space, the objective value of the A* algorithm is set as an upper bound for the total loaded move distance in constraint \ref{eq: upper bound constraint}. 

\begin{equation}
\label{eq: upper bound constraint}
    \displaystyle\sum_{k \in \mathcal{K}}
    \displaystyle\sum_{s \in \mathcal{S}}
    \displaystyle\sum_{s' \in \mathcal{S}}
    \displaystyle\sum_{t \in \mathcal{T}_{s}}
    \displaystyle\sum_{t' \in \mathcal{T}_{s'}}
    z^{k}_{s,t} \cdot y^{k}_{s',t'} \cdot d_{s,s'} \leq c^{UB}
\end{equation}

Given that the number of virtual lanes can be substantial for large warehouse layouts, the symmetry-breaking constraints (27) of CP5 \citep{jimenez2023constraint} are omitted from the model to reduce the model building time. The influence of those constraints is inherently diminished by accounting for the total loaded move distance.

\section{Experiments}
\label{sec:experiments}
The objective of the analysis is to evaluate the performance of the A* algorithm approach (see Section \ref{subsec: Step 2 - Alternative 1: Tree search}) and to compare it with the \gls{cp} approach (see Section \ref{subsec: Step 2 - Alternative 2: Constraint programming}). We refer to the A* algorithm approach as \textit{A*} and the \gls{cp} approach as \textit{CP}.

\subsection{Instance generator and dataset}
We conduct experiments on 3520 instances and publish an even more extensive dataset for future research \citep{pfrommerboemer2024data}. Table \ref{tab:Dataset configurations} shows an overview of the configuration parameters for this study. In all instances, we consider four access directions and one tier. The varied parameters include priority classes, fill percentage, bay layout, and warehouse layout. For each configuration, we create 10 randomly generated instances (seeds 1-10).

\begin{table}[!ht]
\centering
\begin{tabular}{lc}
\toprule
Parameter        & Value      \\
\midrule
Access directions& \{four\} \\
Tiers            & \{1\}\\
Priority classes & \{5, 10\}\\
Fill percentage  & \{40\,\%, 60\,\%, 80\,\%, 90\,\%\} \\
Bay layout       & \{3x3, 4x4, 5x5, 6x6\} \\
Warehouse layout & \{2x2, 3x3, 4x4, 5x5, 6x6, 7x7, 8x8, 9x9, 10x10, 11x11, 12x12\} \\     
\bottomrule
\end{tabular}
\caption{Dataset configurations.}
\label{tab:Dataset configurations}
\end{table}

The test instances consider five or 10 different priority classes. Fill percentages of 40\,\%, 60\,\%, 80\,\%, and 90\% are taken into account. For this study, we only consider quadratic layouts. A 6x6 bay layout describes a bay in which six lanes are arranged next to each other with a depth of six slots. This results in 36 slots per bay. A warehouse layout of 5x5 describes a warehouse that consists of 25 bays that are arranged in a quadratic pattern. Hence, a 6x6 bay layout and a 5x5 warehouse layout combined results in 900 slots. This study includes bay layouts from 3x3 up to 6x6 and warehouse layouts from 2x2 up to 12x12. 
Table \ref{tab:slots per layout configuration} shows the resulting number of slots for each considered bay and warehouse layout configuration. The largest layout configurations include 1296 slots.

\begin{table}[!htbp]
\centering
\begin{tabular}{lrrrrrrrrrrr}
\toprule
 & \multicolumn{11}{c}{Warehouse layout}                                    \\
  \cmidrule{2-12}
 Bay layout & 2x2 & 3x3 & 4x4 & 5x5 & 6x6  & 7x7 & 8x8  & 9x9 & 10x10 & 11x11 & 12x12 \\
           \midrule
3x3        & 36  & 81  & 144 & 225 & 324  & 441 & 576  & 729 & 900   & 1089  & 1296  \\
4x4        & 64  & 144 & 256 & 400 & 576  & 784 & 1024 &     &       &       &       \\
5x5        & 100 & 225 & 400 & 625 & 900  &     &      &     &       &       &       \\
6x6        & 144 & 324 & 576 & 900 & 1296 &     &      &     &       &       &      \\
\bottomrule
\end{tabular}
\caption{Number of slots for each layout configuration.}
\label{tab:slots per layout configuration}
\end{table}

\subsection{Analysis}

We conduct computational experiments on the Linux High-Performance-Computing Cluster LiDo3 at the Technical University of Dortmund. We run all experiments on a node with 4 × Intel Xeon E5 4640v4 2.1 GHz and 512 GB RAM. The algorithms are implemented in Python using Google OR-Tools 9.5 with default parameters for the network flow model and the \gls{cp} model. The solution time of the A* algorithm is limited to 10 minutes. The solution time of the \gls{cp} model is limited to 60 minutes.
We present the findings using examples. For completeness, a detailed overview of the results is provided in the appendix \ref{detailed_results_Astar} and \ref{detailed_results_CP}.

\paragraph{Solved instances}
Table \ref{tab:Solved instances for A*} shows the number of solved instances by A* for each instance configuration. The results show that instances with 10 priority classes are harder to solve than instances with five priority classes. An example is given for a 5x5 bay layout, 6x6 warehouse layout, and 90\,\% fill percentage. For five priority classes, all instances have been solved, for 10 priority classes only four have been solved. 
Further, the bay layout has a stronger impact on the solvability of an instance than the warehouse layout. For example, the instances with a 5x5 bay layout and 6x6 warehouse layout are solvable for the A*, while instances with a 6x6 bay layout and 5x5 warehouse layout are barely solvable for fill percentages over 40\,\%. Both configurations contain the same number of slots.
\par
Also, a larger fill percentage seem to make instances harder to solve up to a certain point. The number of solved instances decreases with a rising fill percentage from 40\,\% to 80\,\%. The comparison of 80\,\% fill percentage and 90\,\% shows that the instances with 80\,\% fill percentage are harder to solve than the ones with 90\,\%. The reason for that is that a high 90\,\% fill percentage heavily restricts the potential slots for relocations. Hence, the A* has to evaluate fewer nodes to find a solution.

\begin{figure}[!htbp]
    \centering
    \includegraphics[width=1\textwidth]{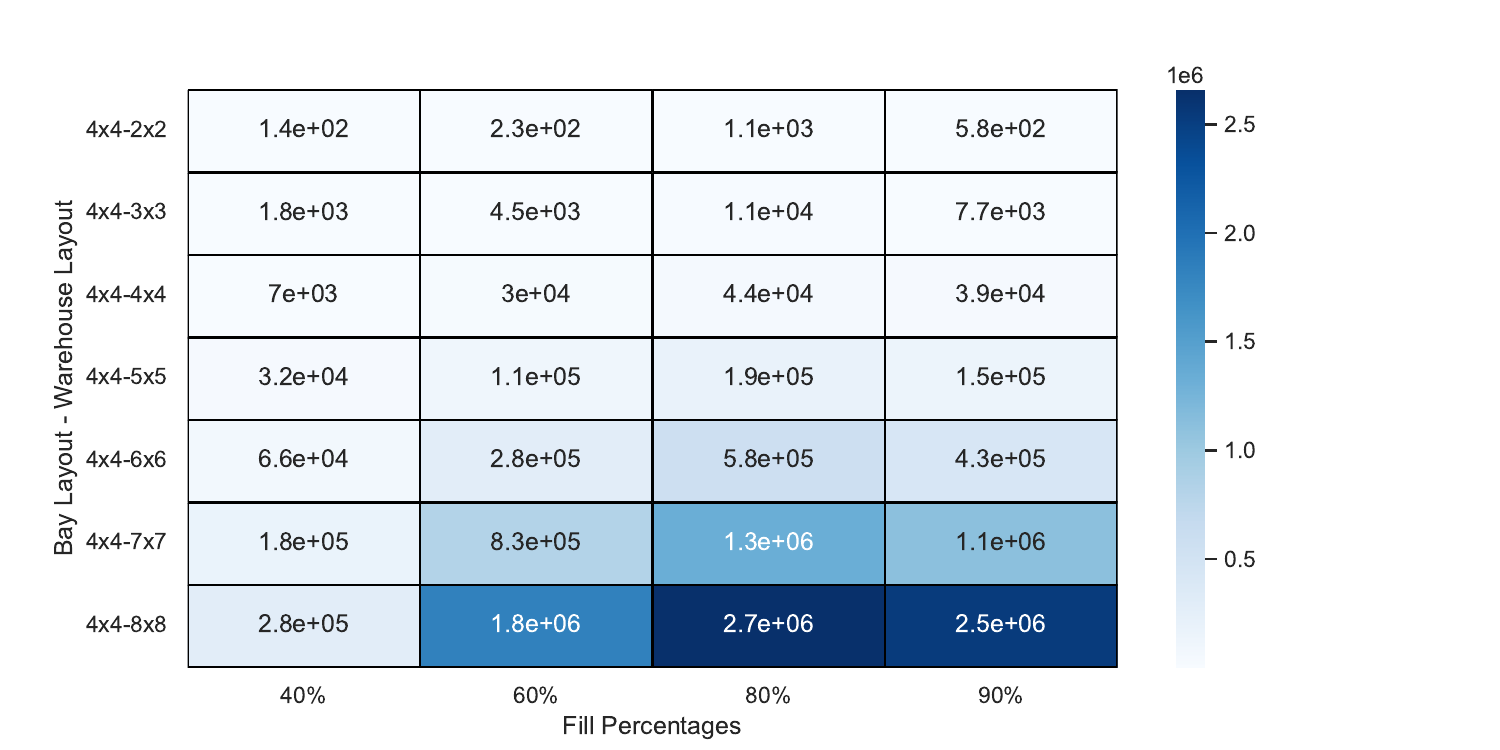}
    \caption{Mean number of evaluated nodes in A* for 10 priority classes.}
    \label{fig:Number of evaluated nodes}
\end{figure}

\begin{table}[!htbp]
\centering
\begin{tabular}{llrrrrlrrrr}
\toprule
  & Priority classes & \multicolumn{4}{l}{5} & & \multicolumn{4}{l}{10} \\
  & Fill percentages & 40\,\% & 60\,\% & 80\,\% & 90\,\%& & 40\,\% & 60\,\% & 80\,\% & 90\,\% \\
  
Bay layout & Warehouse layout &      &     &     &     &   &     &     &     &     \\
\midrule
3x3 & 2x2   &  10 &  10 &  10 &  10 &   &  10 &  10 &  10 &  10 \\
  & 3x3  &  10 &  10 &  10 &  10 &   &  10 &  10 &  10 &  10 \\
  & 4x4  &  10 &  10 &  10 &  10 &   &  10 &  10 &  10 &  10 \\
  & 5x5  &  10 &  10 &  10 &  10 &   &  10 &  10 &  10 &  10 \\
  & 6x6  &  10 &  10 &  10 &  10 &   &  10 &  10 &  10 &  10 \\
  & 7x7  & 10 &  10 &  10 &  10 &   &  10 &  10 &  10 &  10 \\
  & 8x8  &  10 &  10 &  10 &  10 &   &  10 &  10 &  10 &  10 \\
  & 9x9  & 10 &  10 &  10 &  10 &   &  10 &  10 &  10 &  10 \\
  & 10x10 & 10 &  10 &  10 &  10 &   &  10 &  10 &  10 &  10 \\
  & 11x11 &  10 &  10 &  10 &  10 &   &  10 &  10 &  10 &  10 \\
  & 12x12 & 10 &  10 &   \textbf{9} &  10 &   &  10 &  10 &   \textbf{8} &  10 \\
\midrule
4x4 & 2x2  &  10 &  10 &  10 &  10 &   &  10 &  10 &  10 &  10 \\
  & 3x3  &  10 &  10 &  10 &  10 &   &  10 &  10 &  10 &  10 \\
  & 4x4  &  10 &  10 &  10 &  10 &   &  10 &  10 &  10 &  10 \\
  & 5x5  &  10 &  10 &  10 &  10 &   &  10 &  10 &  10 &  10 \\
  & 6x6  &  10 &  10 &  10 &  10 &   &  10 &  10 &  10 &  10 \\
  & 7x7  &  10 &  10 &  10 &  10 &   &  10 &  10 &  10 &  10 \\
  & 8x8  &  10 &  10 &  10 &  10 &   &  10 &   \textbf{9} &   \textbf{3} &  \textbf{9} \\
  \midrule
5x5 & 2x2  &  10 &  10 &  10 &  10 &   &  10 &  10 &  10 &  10 \\
  & 3x3  &  10 &  10 &  10 &  10 &   &  10 &  10 &  10 &  10 \\
  & 4x4  &  10 &  10 &  10 &  10 &   &  10 &  10 &  10 &   9 \\
  & 5x5  &  10 &  10 &  10 &  10 &   &  10 &  10 &  10 &  10 \\
  & 6x6  &  10 &  10 &   \textbf{5} &  10 &   &  10 &  10 &   \textbf{0} &   \textbf{4} \\
  \midrule
6x6 & 2x2  &  10 &  10 &  10 &   \textbf{9} &   &  10 &  10 &  10 &  10 \\
  & 3x3  &  10 &  10 &  10 &   \textbf{7} &   &  10 &  10 &  10 &   \textbf{7} \\
  & 4x4  &  10 &  10 &  10 &   \textbf{6} &   &  10 &  10 &  10 &   \textbf{1} \\
  & 5x5  &  10 &   \textbf{5} &  0  &  0  &   &  10 &   \textbf{2} &   0 &  0  \\
  & 6x6  &  \textbf{3} & 0   &  0  &  0  &   &   \textbf{0} &  0  &  0  &  0  \\
\bottomrule
\end{tabular}
\caption{Solved instances for A*.}
\label{tab:Solved instances for A*}
\end{table}
Figure \ref{fig:Number of evaluated nodes} shows the mean number of evaluated nodes for instances with a 4x4 bay layout. The mean number of evaluated nodes is an indicator of the solution effort of each instance. The mean number of evaluated nodes for 80\,\% fill percentage instances exceeds the number for 90\,\% fill percentage. 
Consequently, the solution effort for 80\,\% is higher than for 90\,\% which results in an A* time-out for smaller layout configurations than for 90\,\% fill percentage. 

\begin{figure}[!htbp]
    \centering
    \includegraphics[width=1\textwidth]{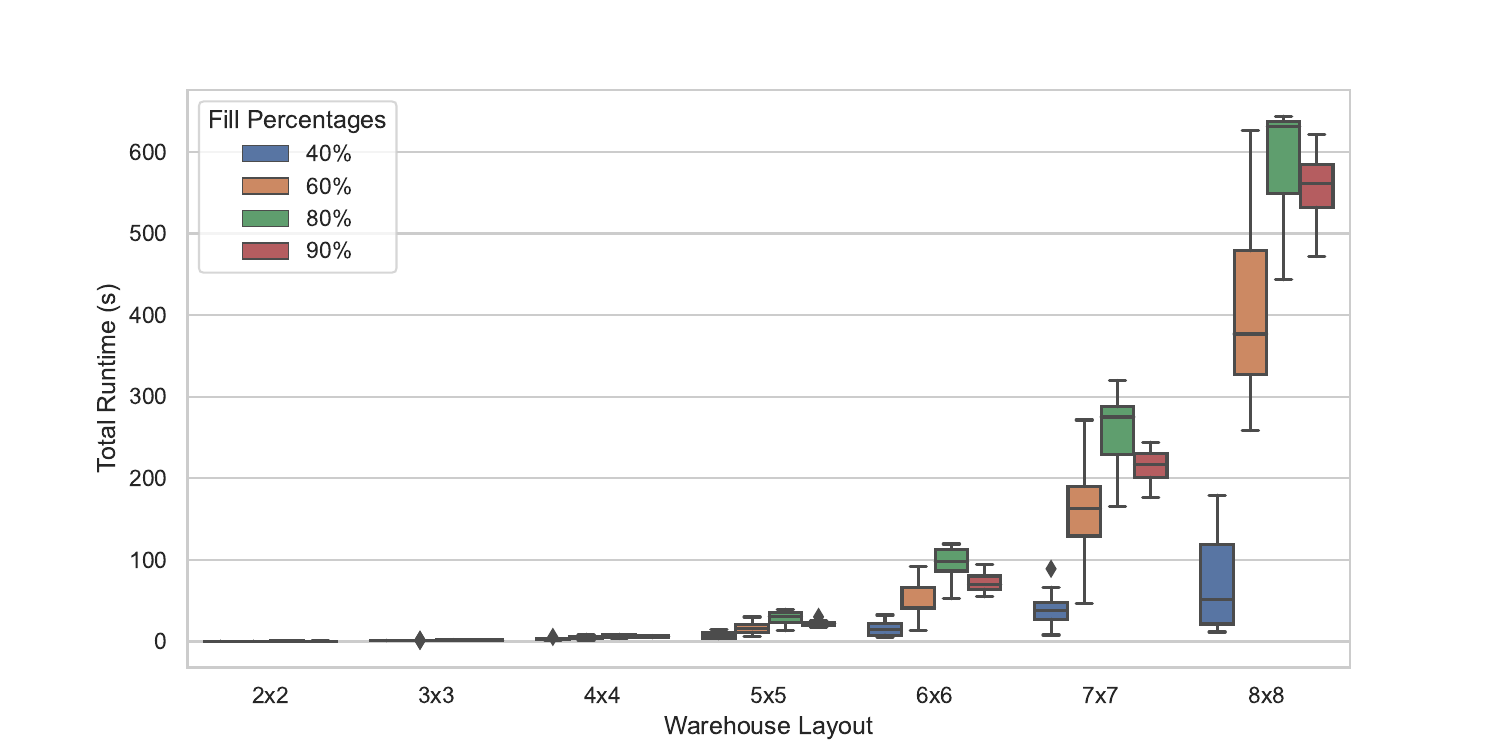}
    \caption{Total runtimes for A* for a 4x4 bay layout and 10 priority classes.}
    \label{fig:total runtime}
\end{figure}

\paragraph{Runtimes}
The total runtimes in Figure \ref{fig:total runtime} include the access direction fixing pre-processing step and the solution time of the A* algorithm itself. The total runtimes increase with the number of evaluated nodes. Consequently, the total runtimes increase with the size of the bay and warehouse layout. Further, the total runtimes increase with the fill percentage until 80\,\%. However, the solving times for the 90\,\% fill percentage instances are lower than for the 80\,\%. The reason for this is that more nodes have to be evaluated for 80\,\% fill percentage than for 90\,\% fill percentage (see Figure \ref{fig:Number of evaluated nodes}).

\begin{figure}[!ht]
    \centering
    \includegraphics[width=1\textwidth]{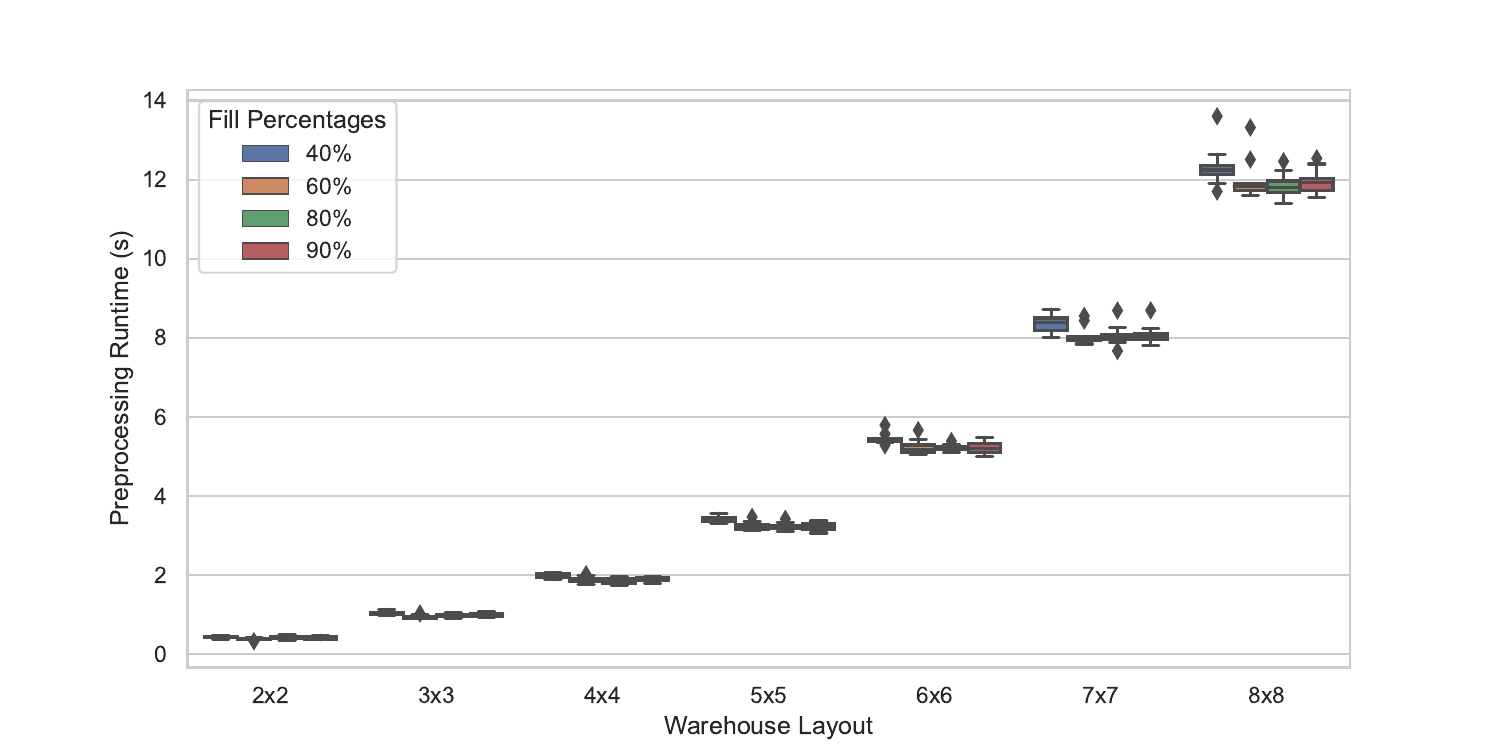}
    \caption{Access direction fixing (pre-processing) runtime for 4x4 bay layout and 10 priority classes.}
    \label{fig: NFM runtime}
\end{figure}

The runtimes for the access direction fixing pre-processing step are shown in Figure \ref{fig: NFM runtime} for the same instances as in Figure \ref{fig:total runtime}. Figure \ref{fig: NFM runtime} shows that the pre-processing runtime is not dependent on the fill percentages, but only on the size of the layout configuration. The pre-processing runtime increases in proportion to the number of bays because the pre-processing step has to be performed for each bay individually. Larger bay layouts cause longer pre-processing times. This becomes evident when comparing two layout configurations with the same number of slots: The mean pre-processing runtime for a 4x4 bay layout and a 5x5 warehouse layout is 3.31\,s for a 5x5 bay layout and a 4x4 warehouse layout the mean is 4.39\,s. 
The number of priority classes does not show a significant impact on the pre-processing time.
The maximum pre-processing time of all solved instances is 39.25\,s for a 6x6 bay layout, 6x6 warehouse layout, 40\,\% fill percentage, and five priority classes. 

\paragraph{Objective value}
To analyze the A* approach, we compare its objective value (total loaded move distance for the minimal number of moves) with the exact \gls{cp} approach. Table \ref{tab: Comparison of optimally solved instances} shows the number of proven optimally solved instances for each solution approach.
Please note that we only run the \gls{cp} model if a solution was found by the A*, because the A* solution serves as an upper bound for the \gls{cp} model.
The number 20 indicates that all instances for that configuration were solved proven optimally. If the number of instances for the A* is lower than the number of optimally solved instances by the \gls{cp} model, a difference between the A* objective value and the proven optimal objective value of the \gls{cp} model is found.
Overall, the \gls{cp} model found the optimal solution in 1364 instances. The A* found the same solution in 97.43\,\% (1329) of these instances. In 35 instances the \gls{cp} model found an optimal solution that differs from the A* solution.
\par
The A* can find the proven optimal solution for almost every 3x3 and 4x4 bay layout configuration that the \gls{cp} model was able to solve to optimality. Only for fill percentages of 90\,\% few suboptimal solutions exist. For 5x5 and 6x6 bay layouts, more instances cannot be optimally solved for 80\,\% and 90\,\% fill percentages by A*. Bay layout and fill percentage have the biggest impact on the optimality of the A* solution. The warehouse layout has little to no impact.

\begin{table}[!ht]
\centering
\begin{tabular}{llrrrrrrrr}
\toprule
  & Fill percentages & \multicolumn{2}{c}{40\,\%} & \multicolumn{2}{c}{60\,\%} & \multicolumn{2}{c}{80\,\%} & \multicolumn{2}{c}{90\,\%} \\
    \cmidrule(lr){3-4} \cmidrule(lr){5-6} \cmidrule(lr){7-8} \cmidrule(lr){9-10}
  & {} &  A* &  CP &  A* &  CP &  A* &  CP &  A* &  CP \\
Bay layout & Warehouse layout &     &     &     &     &     &     &     &     \\
\midrule
3x3 & 2x2  &  20 &  20 &  20 &  20 &  20 &  20 &  \textbf{18} &  \textbf{20} \\
  & 3x3  &  20 &  20 &  20 &  20 &  20 &  20 &  20 &  20 \\
  & 4x4  &  20 &  20 &  20 &  20 &  20 &  20 &  \textbf{19} &  \textbf{20} \\
  & 5x5  &  20 &  20 &  20 &  20 &  20 &  20 &  20 &  20 \\
  & 6x6  &  20 &  20 &  20 &  20 &  20 &  20 & \textbf{18} & \textbf{19} \\
  & 7x7  &  20 &  20 &  20 &  20 &  20 &  20 &  17 &  17 \\
  & 8x8  &  20 &  20 &  20 &  20 &  19 &  19 &   9 &   9 \\
  & 9x9  &  20 &  20 &  20 &  20 &  15 &  15 &   9 &   9 \\
  & 10x10 &  20 &  20 &  20 &  20 &   8 &   8 &   1 &   1 \\
  & 11x11 &  20 &  20 &  20 &  20 &   4 &   4 &   1 &   1 \\
  & 12x12 &  10 &  10 &  10 &  10 &   3 &   3 &   1 &   1 \\
  \midrule
4x4 & 2x2  &  20 &  20 &  20 &  20 &  20 &  20 &  \textbf{18} & \textbf{20} \\
  & 3x3  &  20 &  20 &  20 &  20 &  20 &  20 & \textbf{14} & \textbf{19} \\
  & 4x4  &  20 &  20 &  20 &  20 &  15 &  15 &  \textbf{6} &  \textbf{7} \\
  & 5x5  &  20 &  20 &  18 &  19 &   4 &   4 &  0   &  0   \\
  & 6x6  &  20 &  20 &  17 &  17 &   1 &   1 &   0  &   0  \\
  & 7x7  &  20 &  20 &  12 &  12 &   0  &   0  &  0   &   0  \\
  & 8x8  &  20 &  20 &   9 &   9 &    0 &   0  &   0  &    0 \\
  \midrule
5x5 & 2x2  &  20 &  20 &  \textbf{19} &  \textbf{20} &  \textbf{15} & \textbf{19} &   \textbf{6} &  \textbf{18} \\
  & 3x3  &  20 &  20 &  \textbf{19} &  \textbf{20} &   \textbf{2} &   \textbf{3} &  0  & 0   \\
  & 4x4  &  20 &  20 &   8 &   8 &  0  &  0  &  0  &  0  \\
  & 5x5  &  20 &  20 &   1 &   1 &  0  &  0  &  0  &  0  \\
  & 6x6  &  18 &  18 &   0 &   0 &  0  &  0  &  0  &  0  \\
  \midrule
6x6 & 2x2  &  20 &  20 &  20 &  20 &   \textbf{3} &   \textbf{6} &  0  &  0  \\
  & 3x3  &  20 &  20 &   2 &   2 &  0  &  0  &  0  &  0  \\
  & 4x4  &  15 &  15 &   0 &   0 &  0  &  0  &  0  &  0  \\
  & 5x5  &   5 &   5 &   0 &   0 &     &     &     &     \\
  & 6x6  &   0  &  0 &     &     &     &     &     &     \\
\bottomrule
\end{tabular}
\caption{Number of proven optimal solved instances.}
\label{tab: Comparison of optimally solved instances}
\end{table}

\begin{figure}[!ht]
    \centering
    \includegraphics[width=\textwidth]{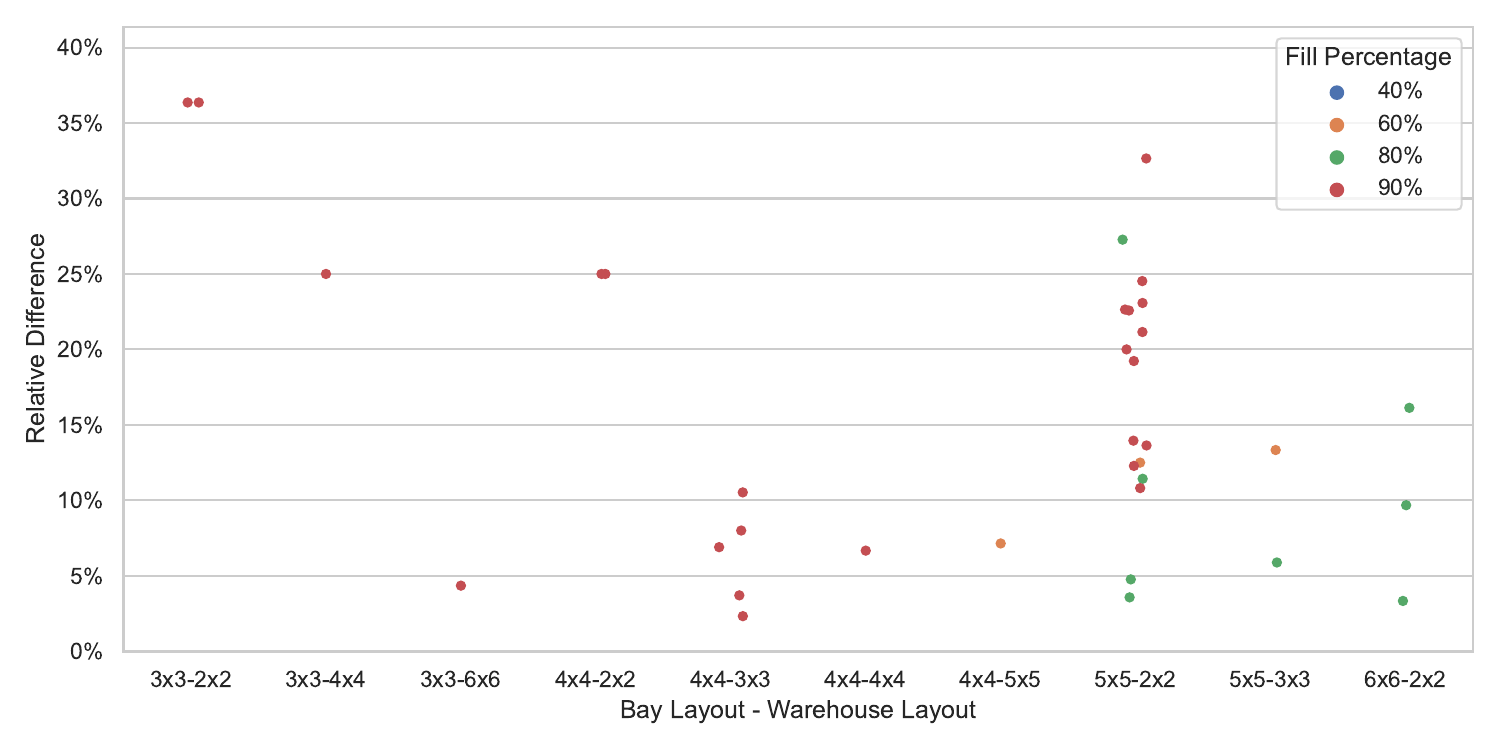}
    \caption{Relative difference between total loaded move distance of the A* and \gls{cp} approach for instances optimally solved by \gls{cp}.}
    \label{fig:relative_difference_Astar_CP}
\end{figure}

Figure \ref{fig:relative_difference_Astar_CP} shows an overview of the relative difference between the total loaded move distance in the A* solution and the \gls{cp} solution if the objective value differs. A value of 20\,\% indicates that the loaded move distance of the A* solution is 20\,\% longer than the optimal loaded move distance found by the \gls{cp} modle. For most layout configurations no or only a few data points exist. For a 5x5 bay layout and a 2x2 warehouse layout, the A* was not able to find the loaded move distance optimal solution in overall 19 instances. For this layout configuration, the mean relative difference of the differing instances is 17.4\,\% across all fill percentage groups. The maximum relative difference for this configuration is 32.6\,\%.

\begin{figure}[!ht]
    \centering
    \includegraphics[width=0.5\textwidth]{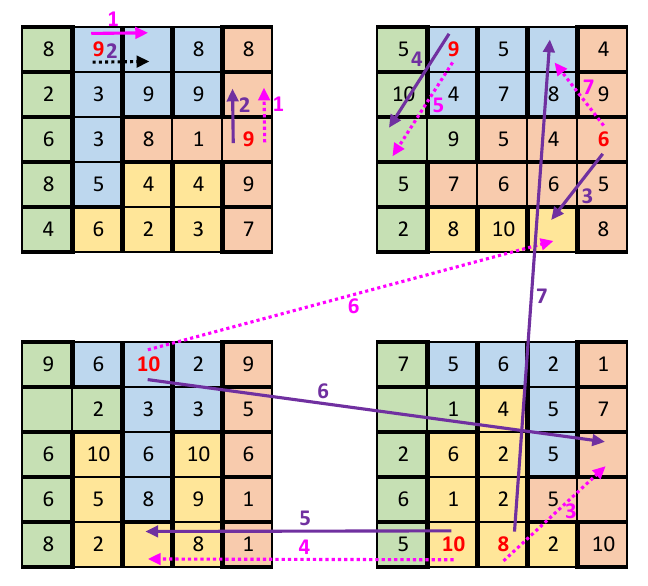}
    \caption[Move sequence of the A* and \gls{cp} approach in comparison]{Move sequence of A* (purple)  and \gls{cp} (pink, dotted) approach in comparison for a 5x5 bay layout, 2x2 warehouse layout, 90\,\% fill percentage and 10 priority classes}
    \label{fig:file_Size_5x5_Layout_2x2_fill_lvl_0.9_seed_5_max_p_10}
\end{figure}

A closer look into the move sequence of both approaches in comparison explains why the A* total loaded move distance is not optimal for some instances. Figure \ref{fig:file_Size_5x5_Layout_2x2_fill_lvl_0.9_seed_5_max_p_10} shows the move sequence of the A* (purple) and \gls{cp} (pink, dotted) approach in comparison for a 5x5 bay layout, 2x2 warehouse layout, 90\,\% fill percentage, and 10 priority classes. Move 1 of the A* is identical to move 2 of the \gls{cp} approach. The same is true for the moves 2/1, 4/5, and 5/4. The longer loaded move distance of the A* results from the fact that dependencies between moves are not taken into account during the search. The A* selects the shortest moves greedily. However, a shorter move earlier in the sequence can cause longer distances later in the sequence. For example, the A* performs a short move 3 that results in a longer distance for move 6 later. A* move 6 causes a greater distance for move 7. The \gls{cp} model respects these dependencies, which results in a lower total move distance for the same number of moves.

\begin{figure}[!ht]
    \centering
    \includegraphics[width=1\textwidth]{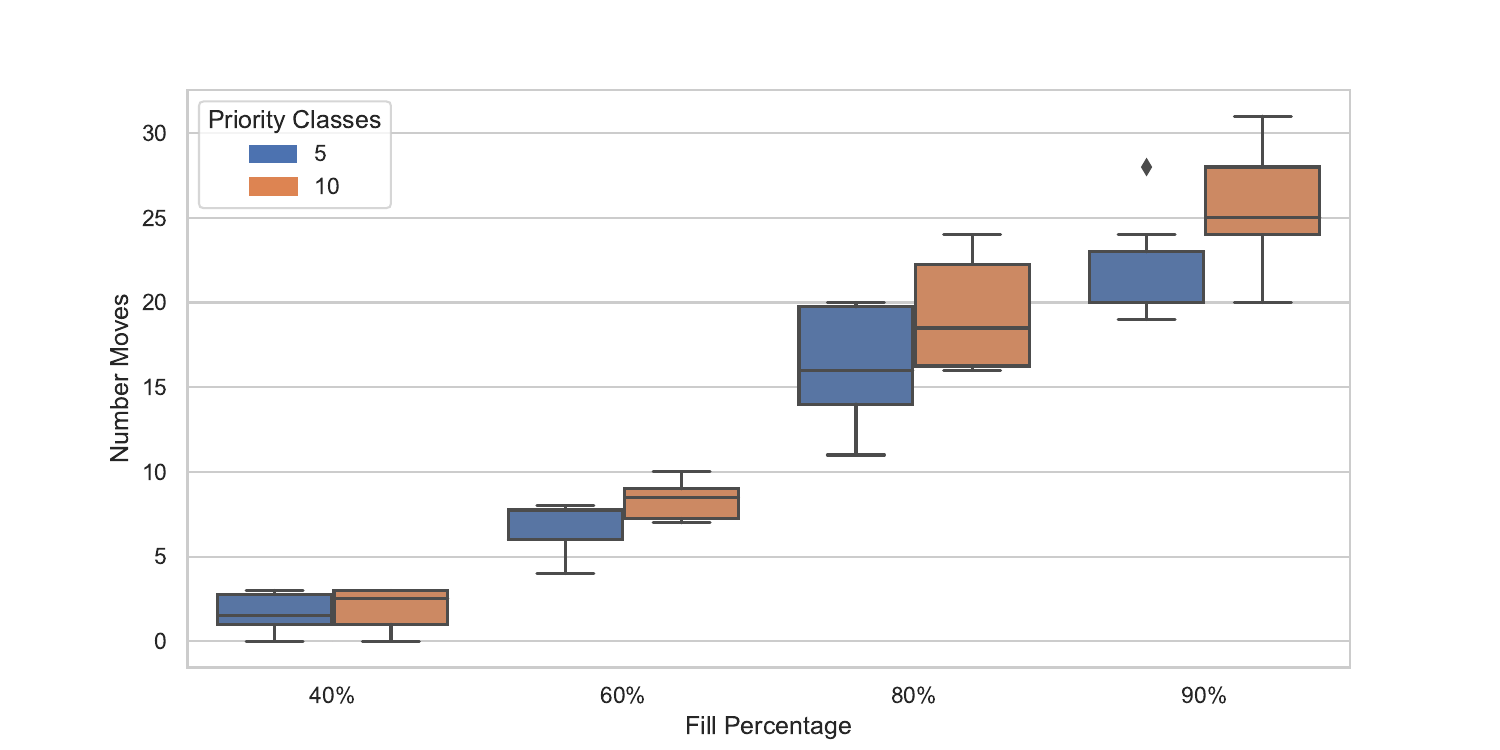}
    \caption[Number of Moves]{Number of moves per fill percentage for each priority class configuration in a 5x5 bay layout and 4x4 warehouse layout.}
    \label{fig:Number of moves in priority classes}
\end{figure}

\paragraph{Move number and loaded move distance}
Figure \ref{fig:Number of moves in priority classes} shows the distribution for the number of moves for five and 10 priority classes in a 5x5 bay layout and 4x4 warehouse layout. For each fill percentage, the number of moves for 10 priority classes is larger than for five priority classes. This is because the chance that a unit load is blocking another unit load increases with the number of priority classes. 

For a better understanding, we calculate the mean likelihood that a unit load is blocking another. We assume that the priority classes are equally distributed. We sum up the likelihood for each priority class that it is blocking if it is standing in front of a random unit load in a virtual lane with a depth of two. Let $\mathcal{P}$ be the set of priority classes and $\Bar{p}$ the maximum priority class $p \in \mathcal{P}$. The mean blockage likelihood $\mathcal{L}_{\Bar{p}}$ is calculated as:

\begin{equation}
\label{eq: upper bound constraint}
    \mathcal{L}_{\Bar{p}} = \frac{1}{\Bar{p}} \sum_{p \in \mathcal{P}} \bigg(1 - \frac{p}{\Bar{p}}\bigg)
\end{equation}
For five priority classes, the mean likelihood of blocking is $\mathcal{L}_{5} =$ 40\,\%. For 10 priority classes, it is $\mathcal{L}_{10} =$ 45\,\%. 
The effect of extra blockages decreases as the number of priority groups increases: $\lim_{{\Bar{p} \to \infty}} \mathcal{L}_{\Bar{p}} = 0.5$.
\par
\begin{figure}[!ht]
    \centering
    \includegraphics[width=1\textwidth]{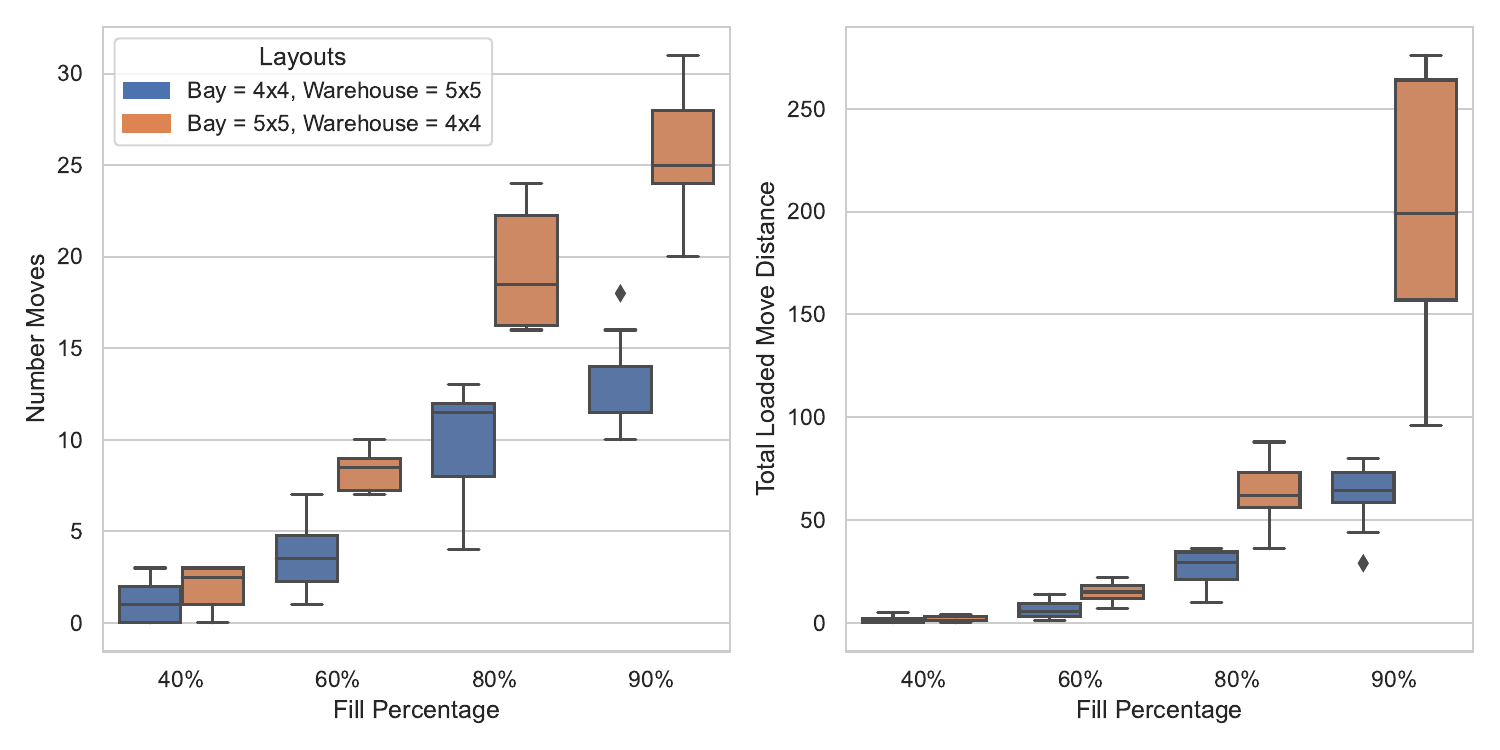}
    \caption[Number of Moves]{Number of moves and total loaded move distance for two layout configurations with the same number of slots and 10 priority classes.}
    \label{fig:Number of Moves layouts}
\end{figure}

Figure \ref{fig:Number of Moves layouts} shows the number of moves and the total loaded move distance for various fill percentages and layout configurations. 
Both configurations have 400 slots in total. The combination of a 4x4 bay layout with a 5x5 warehouse layout requires fewer reshuffling moves than a 5x5 bay layout with a 4x4 warehouse layout. Shallower lanes result in fewer blockages. 
The total loaded move distance shows a similar pattern, as the reshuffling move number and loaded move distance are strongly correlated. Further, the figure displays that a higher fill percentage results in more reshuffling moves and higher total loaded move distances. Especially, the step from 80\,\% to 90\,\% shows a substantial increase in the total loaded move distance.

\begin{figure}[!ht]
    \centering
    \includegraphics[width=1\textwidth]{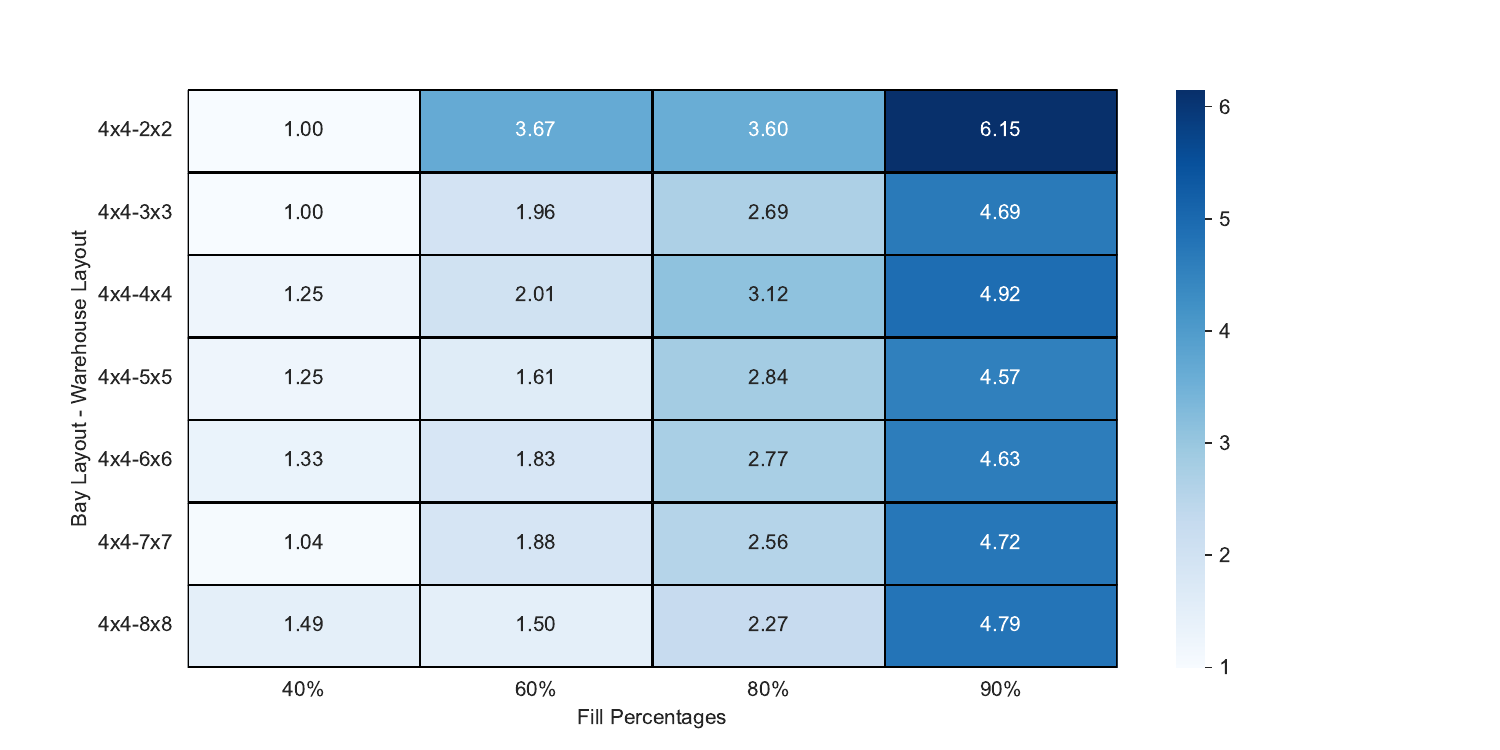}
    \caption{Mean loaded move distance for instances with 10 priority classes.}
    \label{fig: Mean move distance}
\end{figure}

Figure \ref{fig: Mean move distance} presents the mean loaded move distance for instances with a 4x4 bay layout and 10 priority classes. It shows that the mean loaded move distance per move increases with the fill percentage. The reason for this is that fewer repositioning slots are available in close range. Larger warehouse layouts cause smaller mean loaded move distances because the blocking unit loads are less likely at the edge of a warehouse. Hence, more potential repositioning slots are nearby. For a 40\,\% fill percentage, this effect does not show because enough open slots are available in the surrounding area of a blocking unit load.

\section{Conclusion}
\label{sec:conclusion}
In this work, we extended the \gls{upmp} from a single bay to multiple bays. As the objective, we considered the total loaded move distance for the minimal number of moves. To solve the problem, we adapted the two-step approach from \cite{pfrommer2022solving} that is based on a network flow model and an A* tree search procedure. However, given the potential for suboptimal solutions of the tree search in terms of loaded move distance due to inter-move dependencies, 
we devised an alternative strategy proposing an exact \gls{cp} approach. 
In this study, we meticulously evaluated and contrasted the performance of both approaches.
\par
The experiments show that the A* approach can achieve optimal solutions for the total loaded move distance in almost all instances that the \gls{cp} approach was able to solve optimally while providing a tremendously faster runtime. Suboptimal solutions were particularly noticeable in instances characterized by larger bay layouts and high fill levels. At these high fill levels of 80\,\% or 90\,\%, the reduced availability of repositioning options within short distances heightens the significance of optimizing loaded move distances. 
Further, we found that an increase in fill percentage does not necessarily increase the A* solution time, even though more moves are necessary because of a reduced branching factor by a smaller number of possible moves. Overall, the number of reshuffling moves increases with the number of priority classes and the size of the layout configuration. 
The bay size has a significantly larger impact on the number of reshuffling moves than the warehouse size. Hence, it can be beneficial to split the same number of slots into multiple bays. 


\par
Future work should extend the \gls{mupmp} to multiple agents. Hence, dispatching moves to the available \glspl{amr} instead of a single-move sequence.
The objective function for multiple agents could focus on the minimization of total travel time or an evenly distributed workload to minimize the makespan. 
In addition, cases in which a lower move distance can be achieved by additional moves should be discussed. 
Moreover, exploring the integration of the \gls{mupmp} into a simulation model akin to \cite{pfrommer2022slapstack}, which comprehensively addresses pertinent operational decision problems \cite{pfrommer2020autonomously}, would provide valuable insights into the collective effect on warehouse performance.

\newpage
\bibliographystyle{apacite}
\bibliography{Literature.bib}

\appendix
\newpage
\section{Appendix}
\label{appendix}

\subsection{Complete constraint programming model}
\label{appendix_constraint programming}
In this section, the complete \gls{cp} model is presented. Let $\bar s$ denote the number of virtual lanes and $\bar t$ represent the number of positions in each virtual lane $s$. Accordingly, $\mathcal{S} = \{1, . . ., \bar s\}$ is the set of virtual lanes and $\mathcal{T}_{s} = \{1, . . .,\bar t\}$ is the set of positions for virtual lane $s \in \mathcal{S}$.
A slot ($s$,$t$) is defined by virtual lane $s \in \mathcal{S}$ and position $t  \in \mathcal{T}_{s}$.  Let $d_{s, s'}$ be the distance between virtual lane $s \in \mathcal{S}$ and $s' \in \mathcal{S}$.
Let $\bar k$ be the permissible number of moves in the model. $\mathcal{K}$ is the set of stages, defined as $\mathcal{K} = \{1, . . ., \bar k\}$. Stage 0 signifies the initial layout.  $\mathcal{K}^{0} = \{0,1, . . ., \bar k\}$ is the set of stages, including the initial layout. Let $\Bar{p}$ denote the maximum priority class and let $\mathcal{P}$ be the set of priority classes $\mathcal{P} \in \{ 1, ..., \Bar{p} \}$. An empty slot has the priority class 0.
Further, the variables $x^{k}_{s,t}$, $y^{k}_{s,t}$, $\delta^{k}_{s,t}$, $z^{k}_{s,t}$, and $w^{k}_{s,t}$ are introduced.
\begin{align*}
x^{k}_{s,t} &=
    \begin{cases}
      0 & \text{If the slot ($s$,$t$) is empty during stage $k$}\\
      p & \text{If a unit load with priority $p$ is placed in slot ($s$,$t$) during stage $k$}
    \end{cases}
\\
&\hspace*{3.8em} \forall s \in \mathcal{S}, \forall t \in \mathcal{T}_{s}, \forall k \in \mathcal{K}^{0}
\\
y^{k}_{s,t} &=
    \begin{cases}
      1 & \text{If a unit load is moved to slot ($s$,$t$) during stage $k$ } \\
      0 & \text{Otherwise}
    \end{cases}
\\
&\hspace*{3.8em} \forall s \in \mathcal{S}, \forall t \in \mathcal{T}_{s}, \forall k \in \mathcal{K}
\\
\delta^{k}_{s,t} &=
    \begin{cases}
      1 & \text{If there is a unit load in slot ($s$,$t$) is empty during stage $k$}\\
      0 & \text{If a slot ($s$,$t$) is empty during stage $k$} 
    \end{cases}
\\
&\hspace*{3.8em} \forall s \in \mathcal{S}, \forall t \in \mathcal{T}_{s}, \forall k \in \mathcal{K}^{0}
\\
z^{k}_{s,t} &=
    \begin{cases}
      1 & \text{If a unit load is removed from slot ($s$,$t$) during stage $k$}\\
      0 & \text{Otherwise}
    \end{cases}
\\
&\hspace*{3.8em} \forall s \in \mathcal{S}, \forall t \in \mathcal{T}_{s} , \forall k \in \mathcal{K}
\\
w^{k}_{s,t} &=
    \begin{cases}
      1 & \text{If there is a blocking unit load in slot ($s$,$t$) during stage $k$}\\
      0 & \text{Otherwise}
    \end{cases}
\\
&\hspace*{3.8em} \forall s \in \mathcal{S}, \forall t \in \mathcal{T}_{s} , \forall k \in \mathcal{K}
\end{align*}

\begin{flalign}
    & min
    \displaystyle\sum_{k \in \mathcal{K}}
    \displaystyle\sum_{s \in \mathcal{S}}
    \displaystyle\sum_{s' \in \mathcal{S}}
    \displaystyle\sum_{t \in \mathcal{T}_{s}}
    \displaystyle\sum_{t' \in \mathcal{T}_{s'}}
    z^{k}_{s,t} \cdot y^{k}_{s',t'} \cdot d_{s,s'} &
    \label{eq:A1}
    \\
    & \text{Subject to:} & \nonumber
    \\
    & x_{s,t}^0 = f_{s,t} \quad & \forall s \in \mathcal{S}, \forall t \in \mathcal{T}_{s} &
    \label{eq:A2}
    \\
    & |\{x^{k}_{s,t} : s \in \mathcal{S}, t \in \mathcal{T}_{s}, x^{k}_{s,t} = p\}| = m_p \quad & p \in \mathcal{P}^0, k \in \mathcal{K} &
    \label{eq:A3}
    \\
    & \displaystyle\sum_{s \in \mathcal{S}}
    \displaystyle\sum_{t \in \mathcal{T}_{s}}
    y^{k}_{s,t} = 1 \quad & \forall k \in \mathcal{K} &
    \label{eq:A4}
    \\ 
    & x^{k}_{s,t} \leq \bar p \cdot \delta^{k}_{s,t} & \forall s \in \mathcal{S}, \forall t \in \mathcal{T}_{s}, \forall k \in \mathcal{K}^0
    \label{eq:A5}
    \\
    & \delta^{k}_{s,t} \leq x^{k}_{s,t} & \forall s \in \mathcal{S}, \forall t \in \mathcal{T}_{s}, \forall k \in \mathcal{K}^0
    \label{eq:A6}
    \\
    & x^{k-1}_{s,t} \leq x^{k}_{s,t} + \bar p \, (1-\delta^{k}_{s,t}) & \forall s \in \mathcal{S}, \forall t \in \mathcal{T}_{s}, \forall k \in \mathcal{K}
    \label{eq:A7}
    \\
    & x^{k}_{s,t} \leq x^{k-1}_{s,t} + \bar p \, (1-\delta^{k-1}_{s,t}) & \forall s \in \mathcal{S}, \forall t \in \mathcal{T}_{s}, \forall k \in \mathcal{K}
    \label{eq:A8}
    \\
    & y^{k}_{s,1} + \delta^{k-1}_{s,1} \leq 1 & \forall s \in \mathcal{S}, \forall k \in \mathcal{K}
    \label{eq:A9}
    \\
    &  y^{k}_{s,t} \leq \delta^{k+1}_{s,t} & \forall s \in \mathcal{S}, \forall t \in \mathcal{T}, \forall k \in \mathcal{K} \, \backslash \, \{ \Bar{k}  \}
    \label{eq:A10}
    \\
    & \displaystyle\sum_{s \in \mathcal{S}}
    \displaystyle\sum_{t \in \mathcal{T}_{s}}
    z^{k}_{s,t} = 1 \quad & \forall k \in \mathcal{K} &
    \label{eq:A11}
    \\
    & \delta^{k}_{s,t} +z^{k}_{s,t} = y^{k}_{s,t} + \delta^{k-1}_{s,t}  & \forall s \in \mathcal{S}, \forall t \in \mathcal{T}_{s}, \forall k \in \mathcal{K}
    \label{eq:A12}
    \\
    & y^{k}_{s,t+1} + \delta^{k-1}_{s,t+1} + z^{k}_{s,t} \leq \delta^{k-1}_{s,t}  & \forall s \in \mathcal{S}, \forall t \in \mathcal{T}_{s} \, \backslash \, \{ \Bar{t}  \}, \forall k \in \mathcal{K}
    \label{eq:A13}
    \\
    & z^{k+1}_{s,\Bar{t}} + y^{k}_{s,\Bar{t}} \leq \delta^{k}_{s,\Bar{t}} & \forall s \in \mathcal{S}, \forall k \in \mathcal{K} \, \backslash \, \{ \Bar{k}  \}
    \label{eq:A14}
    \\
    & z^{1}_{s,\Bar{t}} \leq \delta^{0}_{s,\Bar{t}} & \forall s \in \mathcal{S}
    \label{eq:A15}
    \\
    & \displaystyle\sum_{t \in \mathcal{T}_{s}} y^{k}_{s,t} + 
    \displaystyle\sum_{t \in \mathcal{T}_{s}} z^{k+1}_{s,t} \leq 1
    & \forall s \in \mathcal{S}, \forall k \in \mathcal{K} \, \backslash \, \{ \Bar{k}  \}
    \label{eq:A16}
    \\
    & x^{k}_{s,t+1} \leq x^{k}_{s,t} + \Bar{p} \cdot w^{k}_{s,t+1}  & \forall s \in \mathcal{S}, \forall t \in \mathcal{T}_{s} \, \backslash \, \{ \Bar{t}  \}, \forall k \in \mathcal{K}
    \label{eq:A17}
    \\
    & w^{k}_{s,t} + \delta^{k}_{s,t+1} \leq w^{k}_{s,t+1} + 1 & \forall s \in \mathcal{S}, \forall t \in \mathcal{T}_{s} \, \backslash \, \{ \Bar{t}  \}, \forall k \in \mathcal{K}
    \label{eq:A18}
    \\
    & w^{k}_{s,t} \leq \delta^{k}_{s,t}  & \forall s \in \mathcal{S}, \forall t \in \mathcal{T}_{s}, \forall k \in \mathcal{K}
    \label{eq:A19}
    \\
    & w^{k}_{s,1} = 0  & \forall s \in \mathcal{S}, \forall k \in \mathcal{K}
    \label{eq:A20}
    \\
    & x^{k}_{s,t} +1 \leq x^{k}_{s,t+1} + (\bar{p} + 1) \cdot (1 - w^{k}_{s,t+1} + w^{k}_{s,t}) & \forall s \in \mathcal{S}, \forall t \in \mathcal{T}_{s} \, \backslash \, \{ \Bar{t}  \}, \forall k \in \mathcal{K}
    \label{eq:A21}
    \\
     & \displaystyle\sum_{s \in \mathcal{S}}
     \displaystyle\sum_{t \in \mathcal{T}_{s}}
     w^{k}_{s,t} +k \leq \Bar{k} & \forall k \in \mathcal{K}
     \label{eq:A22}
     \\
     \label{eq: upper bound constraint}
    &\displaystyle\sum_{k \in \mathcal{K}}
    \displaystyle\sum_{s \in \mathcal{S}}
    \displaystyle\sum_{s' \in \mathcal{S}}
    \displaystyle\sum_{t \in \mathcal{T}_{s}}
    \displaystyle\sum_{t' \in \mathcal{T}_{s'}}
    z^{k}_{s, t} \cdot y^{k}_{s',t'} \cdot d_{s,s'} \leq c^{UB}&
\end{flalign}

The objective function \eqref{eq:A1} minimizes the total loaded move distance.
Constraints \eqref{eq:A2} assign the initial layout to the virtual lanes of the warehouse. 
Constraints \eqref{eq:A3} ensure that the number of unit loads per priority group and the number of empty slots stays consistent for each stage. The number of moves per stage is restricted to one by constraints \eqref{eq:A4}.
Constraints \eqref{eq:A5} and \eqref{eq:A6} link the variables $x^{k}_{s,t}$ and $\delta^{k}_{s,t}$. $\delta^{k}_{s,t}$ is 0 if $x^{k}_{s,t}$ is 0, else $\delta^{k}_{s,t}$ is 1. Unit loads that are not moved are fixed to their position by constraints \eqref{eq:A7} and \eqref{eq:A8}. Constraints \eqref{eq:A9} impose that a unit load can only be moved to a slot that was empty in the previous stage. Constraints \eqref{eq:A10} prevent the unit loads from being moved in two consecutive stages because the two moves could have been performed in one move. This rule is also called 
"transitive move avoidance" \citep{tierneySolvingPremarshallingProblem2017}. 
Constraints \eqref{eq:A11} set the number of unit loads removed from the virtual lanes to 1 for each stage. 
Constrains \eqref{eq:A12} make sure that if a unit load is in a slot. It has been there in the previous stage or was moved there in the current stage. 
\eqref{eq:A13} impose that a unit load can be removed from a certain slot in a stage $k$ only if the respective slot was occupied and the slot in front ($s,t + 1$) was empty during the previous stage.
Constraints \eqref{eq:A14} impose that a unit load can only be moved to a slot ($s,t$) that was empty during the previous stage. It also requires that the slot ($s,t - 1$) was occupied. Constraints \eqref{eq:A15} do the same for the first stage.
Constraints \eqref{eq:A16} impose that when a unit load is moved to a certain virtual lane, no unit load can be removed from that virtual lane in the next stage.
Constraints \eqref{eq:A17} express that a unit load that is in front of another unit load with a higher priority class is a blocking unit load.
Constraints \eqref{eq:A18} require that if there is a blocking unit load on slot ($s,t$) and slot ($s,t + 1$) is occupied, there is also a blocking unit load on slot ($s,t + 1$).
Constraints \eqref{eq:A19} impose that empty slots can not be blocking.
Constraints \eqref{eq:A20} define that the first position of a virtual lane is never blocking because access is always granted.
Constraints \eqref{eq:A21} determine that if there is no blocking unit load in slot ($s,t$) and there is no higher priority unit load in the slot in front of it, then there is no blocking unit load in slot ($s,t + 1$).
Constraints \eqref{eq:A22} impose that the number of blocking unit loads at each stage cannot be greater than the number of blocking unit loads in the following stages.
Constraint \eqref{eq: upper bound constraint} limits the total loaded move distance to the objective value of the A* algorithm.


\newpage
\subsection{Detailed Results for A*}
\label{detailed_results_Astar}
\tiny
\begin{longtable}{llllllllll}
\toprule
Layout & Fill \% &  Priority &  Number     &  Number   & Mean  & Mean   &  Mean (solved) & Mean (all) & Mean (all)  \\ 
 (Bay-Warehouse)       &       &  classes   &  solved &  optimal& moves & distance & total runtime [s]& total runtime [s] & evaluated nodes \\ 
\midrule
\endhead
3x3-2x2 &             40\% &                 5 &             10 &              10 &             0 &                      0 &                         0.19 &                       0.19 &                         0 \\
   3x3-2x2 &             40\% &                10 &             10 &              10 &             0 &                      0 &                         0.19 &                       0.19 &                         0 \\
   3x3-2x2 &             60\% &                 5 &             10 &              10 &           0.1 &                    0.3 &                         0.18 &                       0.18 &                        26.50 \\
   3x3-2x2 &             60\% &                10 &             10 &              10 &           0.1 &                    0.3 &                         0.17 &                       0.17 &                        26.50 \\
   3x3-2x2 &             80\% &                 5 &             10 &              10 &           0.1 &                    0.4 &                         0.18 &                       0.18 &                        18.70 \\
   3x3-2x2 &             80\% &                10 &             10 &              10 &           0.2 &                    0.7 &                         0.17 &                       0.17 &                        36.80 \\
   3x3-2x2 &             90\% &                 5 &             10 &               9 &           0.6 &                    3.3 &                         0.19 &                       0.19 &                        68.90 \\
   3x3-2x2 &             90\% &                10 &             10 &               9 &           0.5 &                    2.9 &                         0.18 &                       0.18 &                        57.10 \\
\bottomrule
\end{longtable}

\newpage
\subsection{Detailed Results for CP}
\label{detailed_results_CP}
\tiny
\begin{longtable}{lllllllllll}
\toprule
 Layout& Fill \% & Priority &  Number     &  Number   & Mean  & Mean   &  \multicolumn{2}{l}{Model build time [s] } &  \multicolumn{2}{l}{Model runtime [s]} \\ 
 (Bay-Warehouse) &      & Classes  &  solved &  optimal& moves & distance & solved  & all & solved & all \\ 
\midrule
\endhead
 3x3-2x2 &             40\% &                 5 &             10 &              10 &             0 &                      0 &                               - &                            - &                                 - &                              - \\
   3x3-2x2 &             40\% &                10 &             10 &              10 &             0 &                      0 &                               - &                            - &                                 - &                              - \\
   3x3-2x2 &             60\% &                 5 &             10 &              10 &           0.1 &                    0.3 &                            0.39 &                         0.39 &                              0.05 &                           0.05 \\
   3x3-2x2 &             60\% &                10 &             10 &              10 &           0.1 &                    0.3 &                            0.45 &                         0.45 &                              0.05 &                           0.05 \\
   3x3-2x2 &             80\% &                 5 &             10 &              10 &           0.1 &                    0.4 &                             0.4 &                          0.4 &                              0.05 &                           0.05 \\
   3x3-2x2 &             80\% &                10 &             10 &              10 &           0.2 &                    0.7 &                            0.46 &                         0.46 &                              0.05 &                           0.05 \\
   3x3-2x2 &             90\% &                 5 &             10 &              10 &           0.6 &                    2.9 &                            0.48 &                         0.48 &                              0.07 &                           0.07 \\
\end{longtable}
\end{document}